\documentclass[proceedings]{JHEP3}
\usepackage{eurosym}
\usepackage{amssymb}
\usepackage{amsfonts}
\usepackage{amsmath,epsfig}
\usepackage{graphicx}

\newbox\mybox

\newcommand\fverb{\setbox\mybox=\hbox\bgroup\verb}
\newcommand\fverbdo{\egroup\medskip\noindent\fbox{\unhbox\mybox}\ }
\newcommand\fverbit{\egroup\item[\fbox{\unhbox\mybox}]}
\conference{Time-dependent Darboux (SUSY) transformations for non-Hermitian q-systems}
\abstract{We propose time-dependent Darboux (supersymmetric) transformations that provide a scheme for the calculation of explicitly time-dependent solvable non-Hermitian partner Hamiltonians. Together with two Hermitian Hamilitonians the latter form a quadruple of Hamiltonians that are related by two time-dependent Dyson equations 
and two intertwining relations in form of a commutative diagram. Our construction is extended to the entire hierarchy of Hamiltonians obtained from time-dependent Darboux-Crum transformations. 
As an alternative approach we also discuss the intertwining relations for Lewis-Riesenfeld invariants for Hermitian as well as non-Hermitian Hamiltonians that reduce the time-dependent 
equations to auxiliary eigenvalue equations. The working of our propsals is discussed for a hierarchy of explicitly time-dependent rational, hyperbolic, Airy function and nonlocal potentials.}

\title{Time-dependent Darboux (supersymmetric) transformations for
non-Hermitian quantum systems}
\author{Julia Cen, Andreas Fring and Thomas Frith \\
Department of Mathematics, City University London,\\
Northampton Square, London EC1V 0HB, UK \\
E-mail: julia.cen.1@city.ac.uk, a.fring@city.ac.uk, thomas.frith@city.ac.uk}

\input{tcilatex}
\begin{document}

\section{Introduction}

Darboux transformations \cite{darboux} are very efficient tools in the study
of exactly or quasi-exactly solvable systems. Formally they map solutions
and coefficient functions of a partial differential equation to new
solutions and a differential equation of similar form with different
coefficient functions. The classic example is a second order differential
equation of Sturm-Liouville type or time-independent Schr\"{o}dinger
equation (TDSE). Since in this context the Darboux transformation relates
two operators that can be identified as isospectral Hamiltonians, this
scenario has been interpreted as the quantum mechanical analogue of
supersymmetry \cite{Witten:1981nf,Cooper,bagrov1}. Many potentials with
direct physical applications may be generated with this technique, such as
for instance complex crystals with invisible defects \cite%
{MatLongo,correa2015p}. By relating quantum mechanical systems to soliton
solutions of nonlinear differential equations, such as for instance the
Korteweg-de Vries equation, the sine-Gordon equation or the nonlinear Schr%
\"{o}dinger equation, Darboux transformations have also been very
efficiently utilized in the construction of multi-soliton solutions \cite%
{matveevdarboux,correahidden,CCFsineG,guilarte2017perfectly,cen2018asymptotic}%
.

Initially Darboux transformations were developed for stationary equations,
so that the treatment of the full TDSE was not possible. Evidently the
latter is a much more intricate problem to solve, especially for
non-autonomous Hamiltonians. Explicitly time-dependent Darboux
transformations for TDSE, rather than the time-independent Schr\"{o}dinger
equation, were first introduced by Bagrov and Samsonov \cite{bagrov2} and
subsequently generalized to other types of time-dependent systems \cite%
{song2003,suzko2009darboux}. The limitations of the generalization from the
time-independent to the time-dependent Schr\"{o}dinger equation were that
the solutions considered in \cite{bagrov2} force the Hamiltonians involved
to be Hermitian. One of the central purposes of this manuscript is to
overcome this shortcoming and propose fully time-dependent Darboux
transformations that deal directly with the TDSE involving non-Hermitian
Hamiltonians. We extend our analysis to the entire hierarchy of solvable
time-dependent Hamiltonians constructed from generalized versions of
Darboux-Crum transformations. As an alternative scheme we also discuss the
intertwining relations for Lewis-Riesenfeld invariants for Hermitian as well
as non-Hermitian Hamiltonians. These quantities are constructed as auxiliary
objects to convert the fully TDSE into an eigenvalue equation that is easier
to solve and subsequently allows to tackle the TDSE. The class of
non-Hermitian Hamiltonians we consider here is the one of $\mathcal{PT}$%
-symmetric/quasi-Hermitian ones \cite{Urubu,Benderrev,Alirev} that are
related to a Hermitian counterpart by means of the time-dependent Dyson
equation (TDDE) \cite%
{CA,time1,time6,time7,fringmoussa,AndTom1,AndTom2,AndTom3,AndTom4,mostafazadeh2018energy,AndTom5}%
.

Given the interrelations of the various quantities in the proposed scheme
one may freely choose different initial starting points. A quadruple of
Hamiltonians, two Hermitian and two non-Hermitian ones, is related by two
TDDE and two intertwining relations in form of a commutative diagram. This
allows to compute all four Hamiltonians by solving either two intertwining
relations and one TDDE or one intertwining relations and two TDDE, with the
remaining relation being satisfied by the closure of the commutative
diagram. We discuss the working of our proposal by taking two concrete
non-Hermitian systems as our starting points, the Gordon-Volkov Hamiltonian
with a complex electric field and a reduced version of the Swanson model.
From the various solutions to the TDSE we construct explicitly
time-dependent rational, hyperbolic, Airy function and nonlocal potentials.

Our manuscript is organized as follows: In section 2 we review the
time-dependent Darboux transformations for Hermitian Hamiltonians and stress
the limitations of previous results. We propose a new scheme that allows for
the treatment of non-Hermitian Hamiltonians. Subsequently we extend the
Darboux transformations to Darboux-Crum transformations, that is we
construct two hierarchies from intertwining operators build from solutions
previously ignored. In section 3 we discuss the intertwining relations for
Lewis-Riesenfeld invariants. Taking a complex Gordon-Volkov Hamiltonian as
starting point we discuss in section 4 various options of how to close the
commutative diagrams constructing the intertwining operators from different
types of solutions for rational, hyperbolic, Airy function potentials. In
section 5 we start from a reduced version of the Swanson model and carry out
the analysis for two different Dyson maps. In addition we discuss
intertwining relations for Lewis-Riesenfeld invariants for this concrete
system. The solutions to the TDSE discussed in this section depend on the
solutions of an auxiliary equation known as the dissipative Ermakov-Pinney
equation. We discuss in appendix A how to obtain explicit solutions to this
nonlinear second order differential equation. Our conclusions are stated in
section 6.

\section{Time-dependent Darboux-Crum transformations}

\subsection{Time-dependent Darboux transformations for Hermitian systems}

Before introducing the time-dependent Darboux transformations for
non-Hermitian systems we briefly recall the construction for the Hermitian
setting. This revision will not only establish our notation, but it also
serves to highlight why previous suggestions are limited to the treatment of
Hermitian systems. Here we wish to overcome this shortcoming and extend the
theory of Darboux transformations to include the treatment of time-dependent
non-Hermitian Hamiltonians. Our main emphasis is on non-Hermitian systems
that belong to the class of $\mathcal{PT}$-symmetric Hamiltonians, i.e. they
remain invariant under the antilinear transformation $\mathcal{PT}:$ $%
x\rightarrow -x$, $p\rightarrow p$, $i\rightarrow -i$. Such type of systems
are of physical interest as potentially they possess energy operators with
real instantaneous eigenvalues, that are different from the Hamiltonians in
the non-Hermitian case.

The time-dependent Hermitian intertwining relation introduced in \cite%
{bagrov2} reads 
\begin{equation}
\ell \left( i\partial _{t}-h_{0}\right) =\left( i\partial _{t}-h_{1}\right)
\ell ,  \label{HI}
\end{equation}%
where the Hermitian Hamiltonians $h_{0}$ and $h_{1}$ involve explicitly
time-dependent potentials $v_{j}\left( x,t\right) $ 
\begin{equation}
h_{j}\left( x,t\right) =p^{2}+v_{j}\left( x,t\right) ,\qquad j=0,1.
\label{HamiltonianForm}
\end{equation}%
The intertwining operator $\ell $ is taken to be a first order differential
operator 
\begin{equation}
\ell \left( x,t\right) =\ell _{0}\left( x,t\right) +\ell _{1}\left(
x,t\right) \partial _{x}.  \label{ll}
\end{equation}%
In general we denote by $\phi _{j}$, $j=0,1$, the solutions to the two
partner TDSEs $i\partial _{t}\phi _{j}=h_{j}\phi _{j}$. Throughout our
manuscript we use the convention $\hbar =1$. Taking a specific solution $%
u(x,t):=\phi _{0}(x,t)$ to one of these equations, the constraints imposed
by the intertwining relation (\ref{HI}) can be solved by%
\begin{equation}
\ell _{1}\left( x,t\right) =\ell _{1}\left( t\right) ,\quad \ell _{0}\left(
x,t\right) =-\ell _{1}\frac{u_{x}}{u},\quad v_{1}=v_{0}+i\frac{\left( \ell
_{1}\right) _{t}}{\ell _{1}}+2\left( \frac{u_{x}}{u}\right) ^{2}-2\frac{%
u_{xx}}{u},  \label{v1}
\end{equation}%
where, as indicated, $\ell _{1}$ must be an arbitrary function of $t$ only.
At this point the new potential $v_{1}$ might still be complex. However,
besides mapping the coefficient functions, the main practical purpose of the
Darboux transformations is that one also obtains exact solutions $\phi _{1}$
for the partner TDSE $i\partial _{t}\phi _{1}=h_{1}\phi _{1}$ by employing
the intertwining operator. In this case the direct application, that is
acting with (\ref{HI}) on $u$, yields just the trivial solution $\phi
_{1}=lu=0$. For this reason different types of nontrivial solutions were
proposed in \cite{bagrov2} 
\begin{equation}
\hat{\phi}_{1}=\frac{1}{\ell _{1}u^{\ast }},\qquad \text{and\qquad }\tilde{%
\phi}_{1}=\hat{\phi}_{1}\int^{x}\left\vert u\right\vert ^{2}dx^{\prime },
\label{ntsol}
\end{equation}%
which require, however, that one imposes%
\begin{equation}
\ell _{1}(t)=\exp \left[ -\int^{t}\func{Im}\left( v_{0}+2\left( \frac{u_{x}}{%
u}\right) ^{2}-2\frac{u_{xx}}{u}\right) dt^{\prime }\right] .  \label{h1}
\end{equation}%
It is this assumption on the particular form of the solution that forces the
new potentials in the proposal of \cite{bagrov2} to be real $v_{1}=\func{Re}%
\left( v_{0}+2\left( u_{x}/u\right) ^{2}-2u_{xx}/u\right) $. Notice that one
might not be able to satisfy (\ref{h1}), as the right hand side must be
independent of $x$. If the latter is not the case, the solutions in (\ref%
{ntsol}) and the partner Hamiltonian $h_{1}$ do not exist.

Here we also identify another type of nontrivial solutions. Acting with
equation (\ref{HI}) to the right on a solution of the TDSE $i\partial
_{t}\phi _{0}=h_{0}\phi _{0}$, say $\phi _{0}=\tilde{u}$, that is linearly
independent from $\phi _{0}=u$ used in the construction of the intertwining
operator will in general lead to nontrivial solutions 
\begin{equation}
\phi _{1}=\mathcal{L}\left[ u\right] (\tilde{u}),\quad \text{with \ }%
\mathcal{L}\left[ u\right] (f):=\ell _{1}\left( t\right) \left( \partial
_{x}f-\frac{u_{x}}{u}f\right)  \label{second}
\end{equation}%
to the second TDSE $i\partial _{t}\phi _{1}=h_{0}\phi _{1}$. This type of
solution and was overlooked in \cite{bagrov2} and in principle might lead to
complex potentials $v_{1}$ as it is not restricted by any additional
constraints.

\subsection{Time-dependent Darboux transformations for non-Hermitian systems}

In order to extend the previous analysis in the way that allows for other
types of complex potentials, and especially general non-Hermitian
Hamiltonians that are $\mathcal{PT}$-symmetric/quasi-Hermitian \cite%
{Urubu,Benderrev,Alirev}, we make use of the time-dependent Dyson equation
(TDDE) \cite%
{CA,time1,time6,time7,fringmoussa,AndTom1,AndTom2,AndTom3,AndTom4,mostafazadeh2018energy,AndTom5}
for both time-dependent Hermitian Hamiltonians $h_{0}(t)$, $h_{1}(t)$ and
the time-dependent non-Hermitian Hamiltonians $H_{0}(t)$, $H_{1}(t)$ 
\begin{equation}
h_{j}=\eta _{j}H_{j}\eta _{j}^{-1}+i\partial _{t}\eta _{j}\eta
_{j}^{-1},\qquad j=0,1.  \label{Dysoneq}
\end{equation}%
The time-dependent Dyson maps $\eta _{j}(t)$ relate the solutions of the
TDSE $i\partial _{t}\psi _{j}=H_{j}\psi _{j}$ to the previous ones for $\phi
_{j}$ as%
\begin{equation}
\phi _{j}=\eta _{j}\psi _{j},\qquad j=0,1.
\end{equation}%
Using (\ref{Dysoneq}) in the intertwining relation (\ref{HI}) yields 
\begin{equation}
\ell \left( i\partial _{t}-\eta _{0}H_{0}\eta _{0}^{-1}-i\partial _{t}\eta
_{0}\eta _{0}^{-1}\right) =\left( i\partial _{t}-\eta _{1}H_{1}\eta
_{1}^{-1}-i\partial _{t}\eta _{1}\eta _{1}^{-1}\right) \ell .  \label{aux}
\end{equation}%
Multiplying (\ref{aux}) from the left by $\eta _{1}^{-1}$ and acting to the
right on $\eta _{0}f$, with $f(x,t)$ being some arbitrary test function, we
obtain 
\begin{eqnarray}
&&\eta _{1}^{-1}\ell \left[ i\left( \partial _{t}\eta _{0}\right) f+i\eta
_{0}\partial _{t}f-\eta _{0}H_{0}f-i\left( \partial _{t}\eta _{0}\right) f%
\right] =  \notag \\
&&\left( i\eta _{1}^{-1}\ell \eta _{0}\partial _{t}f+i\eta _{1}^{-1}\left(
\partial _{t}\ell \eta _{0}\right) f-H_{1}\eta _{1}^{-1}\ell \eta
_{0}f-i\eta _{1}^{-1}\left( \partial _{t}\eta _{1}\right) \eta _{1}^{-1}\ell
\eta _{0}f\right) .
\end{eqnarray}%
Rearranging the time derivative terms and removing the test function, we
derive the new intertwining relation for non-Hermitian Hamiltonians%
\begin{equation}
L\left( i\partial _{t}-H_{0}\right) =\left( i\partial _{t}-H_{1}\right) L,
\label{IH}
\end{equation}%
where we introduced the new intertwining operator 
\begin{equation}
L:=\eta _{1}^{-1}\ell \eta _{0}.  \label{DarbouxGeneral}
\end{equation}%
We note that $H_{j}-p^{2}$ is in general not only no longer real and might
also include a dependence on the momenta, i.e. $H_{j}$ does not have to be a
potential Hamiltonian and could be nonlocal. Denoting by $\psi _{0}=U=\eta
_{0}^{-1}u~$a particular solution to the TDSE for $H_{0}$, the standard new
solution $\psi _{1}=LU=\eta _{1}^{-1}\ell \eta _{0}\eta _{0}^{-1}u$ remains
trivial. The nontrivial solutions (\ref{ntsol}) generalize to 
\begin{equation}
\hat{\psi}_{1}=\eta _{1}^{-1}\frac{1}{\ell _{1}\left( \eta _{0}U\right)
^{\ast }},\qquad \text{and\qquad }\tilde{\psi}_{1}=\hat{\psi}%
_{1}\int^{x}\left\vert \eta _{0}U\right\vert ^{2}dx^{\prime }.  \label{ntP}
\end{equation}%
The nontrivial solution (\ref{second}) becomes%
\begin{equation}
\psi _{1}=L\left[ U\right] \left( \tilde{U}\right)
\end{equation}%
in the non-Hermitian case. In summary, our quadruple of Hamiltonians is
related as depicted in the commutative diagram%
\begin{equation}
\begin{array}{ccccc}
& H_{0} & ~~\underrightarrow{\eta _{0}}~~ & h_{0} &  \\ 
\eta _{1}^{-1}\mathcal{L}\left[ u\right] \eta _{0} & \downarrow &  & 
\downarrow & \mathcal{L}\left[ u\right] \\ 
& H_{1} & \underrightarrow{\eta _{1}} & h_{1} & 
\end{array}
\label{DH}
\end{equation}

One may of course also try to solve the intertwining relation (\ref{IH})
directly and build the intertwining operator $L$ from a solution $U=\eta
_{0}^{-1}u$ for the TDSE for $H_{0}$ and ignore initially the fact that the
Hamiltonians $H_{0}$ and $H_{1}$ involved are non-Hermitian. To make sense
of these Hamiltonians one still needs to construct the Dyson maps $\eta _{0}$
and $\eta _{1}$. Considering the diagram 
\begin{equation}
\begin{array}{ccccc}
& H_{0} & ~~\underrightarrow{\eta _{0}}~~ & h_{0} &  \\ 
\mathcal{L}\left[ U\right] & \downarrow &  & \downarrow & \mathcal{L}\left[ u%
\right] \\ 
& H_{1} & \underrightarrow{?} & h_{1} & 
\end{array}
\label{dia}
\end{equation}%
in which the TDDE has been solved for $\eta _{0}$, $H_{0}$, $h_{0}$ and $%
H_{1}$, $h_{1}$ have been constructed with intertwining operators build from
the solutions of the respective TDSE, we address the question of whether it
is possible to close the diagram, that is making it commutative. For this to
be possible we require%
\begin{equation}
\mathcal{L}\left[ U\right] =\eta _{1}^{-1}\mathcal{L}\left[ u\right] \eta
_{0}  \label{LLL}
\end{equation}%
to be satisfied. It is easy to verify that (\ref{LLL}) holds if and only if%
\begin{equation}
\eta _{1}\mathcal{=}\eta _{0},\qquad \text{and\qquad }\frac{\eta
_{0}^{-1}u_{x}}{\eta _{0}^{-1}u}=\eta _{0}^{-1}\frac{u_{x}}{u}\eta _{0}.
\label{sc}
\end{equation}%
A solution for the second equation in (\ref{sc}) is for instance $\eta
_{0}=f(x)T_{\alpha }(x)$, with $T_{\alpha }=e^{i\alpha p}$ being a standard
shift operator, i.e. $T_{\alpha }g(x)=g(x+\alpha )$, and $f(x)$ an arbitrary 
$x$-dependent function.

\subsection{Time-dependent Darboux-Crum transformations for Hermitian systems%
}

Next we demonstrate that the iteration procedure of the Darboux
transformation, usually referred to as Darboux-Crum (DC) transformations 
\cite{darboux,crum,matveevdarboux}, will lead also in the time-dependent
case to an entire hierarchy of exactly solvable time-dependent Hamiltonians $%
h_{0}$, $h_{1}$, $h_{2}$, \ldots\ for the TDSEs $i\partial _{t}\phi
^{(n)}=h_{n}\phi ^{(n)}$ related to each other by intertwining operators $%
\ell ^{(n)}$\ 
\begin{equation}
\ell ^{(n)}\left( i\partial _{t}-h_{n-1}\right) =\left( i\partial
_{t}-h_{n}\right) \ell ^{(n)},\qquad n=1,2,\ldots  \label{iter}
\end{equation}%
For $n=1$ this is equation (\ref{HI}) with $\ell =\ell ^{(1)}$ and solutions 
$\phi _{0}=\phi ^{(0)}$, $\phi _{1}=\phi ^{(1)}$. Taking a particular
solution $\phi _{0}=u$ to depend on some parameter $\gamma $, continuously
or discretely, we denote the solutions at different values as $%
u_{i}:=u(\gamma _{i})$. Given now $\ell ^{(1)}=\mathcal{L}\left[ u_{0}\right]
$ from (\ref{ll}) we act with (\ref{iter}) for $n=1$ on $u_{1}$, so that we
can cast the intertwining operator and the solution (\ref{second}) in the
form 
\begin{equation}
\ell ^{(1)}(f)=\mathcal{L}\left[ u_{0}\right] (f)=\ell _{1}\left( t\right) 
\frac{W_{2}[u_{0},f]}{W_{1}[u_{0}]},\quad \qquad \phi ^{(1)}=\ell
^{(1)}(u_{1})=\mathcal{L}\left[ u_{0}\right] (u_{1}),
\end{equation}%
with corresponding time-dependent Hamiltonian%
\begin{equation}
h_{1}=h_{0}-2\left[ \ln W_{1}(u_{0})\right] _{xx}+i\left[ \ln \ell _{1}%
\right] _{t}.
\end{equation}%
We employed here the Wronskian $W_{n}[u_{1},u_{2},\ldots ,u_{n}]:=$ $\det
\omega $ with $\omega _{jk}=\partial ^{j-1}u_{k}/\partial x^{j-1}$ for $%
j,k=1,\ldots ,n$, e.g. $W_{1}[u_{0}]=$ $u_{0}$, $W_{2}[u_{0},u_{1}]=$ $%
u_{0}\left( u_{1}\right) _{x}-u_{1}\left( u_{0}\right) _{x}$, etc., which
allows to write the expressions for the intertwining operator and
Hamiltonians in the hierarchy in a very compact form. Iterating these
equations we obtain the compact closed form for the intertwining operator 
\begin{eqnarray}
\ell ^{(n)}(f) &=&\mathcal{L}\left[ \ell ^{(n-1)}(u_{n-1})\right] (\ell
^{(n-1)}(f))  \label{DCa} \\
&=&\ell _{1}^{n}\left( t\right) \frac{W_{n+1}[u_{0},u_{1},\ldots ,u_{n-1},f]%
}{W_{n}[u_{0},u_{1},\ldots ,u_{n-1}]} \\
&=&\ell _{1}^{n}\left( t\right) \left\vert \Omega \right\vert _{(n+1)(n+1)},
\end{eqnarray}%
where $\left\vert \Omega \right\vert _{(n+1)(n+1)}$ denotes a
quasideterminant \cite{gelfand2005} for the (n+1)$\times $(n+1)-matrix $%
\Omega $ with $\Omega _{jk}=\partial ^{j-1}u_{k}/\partial x^{j-1}$, $\Omega
_{j(n+1)}=\partial ^{j-1}f/\partial x^{j-1}$ for $j=1,\ldots ,n+1$, $%
k=1,\ldots ,n$. For the time-dependent Hamiltonians we derive%
\begin{equation}
h_{n}=h_{0}-2\left[ \ln W_{n}\left( u_{0},u_{1},\ldots ,u_{n-1}\right) %
\right] _{xx}+in\left( \ln \ell _{1}\right) _{t}.  \label{hnn}
\end{equation}%
Nontrivial solutions of the type (\ref{second}) to the related TDSE $%
i\partial _{t}\phi ^{(n)}=h_{n}\phi ^{(n)}$ are then obtained as%
\begin{equation}
\phi ^{(n)}=\ell ^{(n)}(u_{n})\text{.}  \label{nsol}
\end{equation}%
Instead of using the same solution $u_{i}$ of the TDSE for $h_{0}$ at
different parameter values in the closed expression, it is also possible to
replace some of the solutions $u_{i}$ by the second linear independent
solutions $\tilde{u}_{i}$ at the same parameter values, see e.g. \cite%
{CorreaFring,CCFsineG,CenFringHir} and references therein for details. This
choice allows for the treatment of degenerate solutions. Closed expressions
for DC-transformation build from the solutions (\ref{ntP}) can be found in 
\cite{bagrov2}. Below we will illustrate the working of the formulae in this
section with concrete examples.

\subsection{Time-dependent DC transformations for non-Hermitian systems}

The iteration procedure for the non-Hermitian system goes along the same
lines as for the Hermitian case, albeit with different intertwining
operators $L$. The iterated systems are\ 
\begin{equation}
L^{(n)}\left( i\partial _{t}-H_{n-1}\right) =\left( i\partial
_{t}-H_{n}\right) L^{(n)},\qquad n=1,2,\ldots  \label{H1}
\end{equation}%
The intertwining operators read in this case 
\begin{equation}
L^{(n)}(f)=\mathcal{L}\left[ L^{(n-1)}(U_{n-1})\right] (L^{(n-1)}(f))=\ell
_{1}^{n}\left( t\right) \frac{W_{n+1}[U_{0},U_{1},\ldots ,U_{n-1},f]}{%
W_{n}[U_{0},U_{1},\ldots ,U_{n-1}]},
\end{equation}%
and the time-dependent Hamiltonians are%
\begin{equation}
H_{n}=H_{0}-2\left[ \ln W_{n}[U_{0},U_{1},\ldots ,U_{n-1}]\right] _{xx}+in%
\left[ \ln \ell _{1}\right] _{t}.
\end{equation}%
The nontrivial solutions to the related TDSE are then obtained as%
\begin{equation}
\psi ^{(n)}=L^{(n)}(U_{n}).  \label{H4}
\end{equation}%
Notice that in (\ref{H1})-(\ref{H4}) the only Dyson maps involved are $\eta
_{0}$ and $\eta _{1}$. Alternatively we can also express $L^{(n)}=\eta
_{n}^{-1}l^{(n)}\eta _{n-1}$ and $\psi ^{(n)}=\eta _{n}^{-1}\phi ^{(n)}$,
but the computation of the $\eta _{n}$ for $n>1$ is not needed. Since the
solutions (\ref{ntP}) require the Hamiltonians involved to be Hermitian,
hierarchies build on them do not exist in the non-Hermitian case.

\section{Intertwining relations for Lewis-Riesenfeld invariants}

As previously argued \cite{maamache2017pseudo,AndTom4,AndTom5}, the most
efficient way to solve the TDDE (\ref{Dysoneq}), as well as the TDSE, is to
employ the Lewis-Riesenfeld invariants \cite{Lewis69}. The steps in this
approach consists of first solving the evolution equation for the invariants
of the Hermitian and non-Hermitian system separately and subsequently
constructing a similarity transformation between the two invariants. By
construction the map facilitating this transformation is the Dyson map
satisfying the TDDE.

Here we need to find four time-dependent invariants $I_{j}^{h}(t)$ and $%
I_{j}^{H}(t)$, $j=0,1$, that solve the equations%
\begin{equation}
\partial _{t}I_{j}^{H}(t)=i\left[ I_{j}^{H}(t),H_{j}(t)\right] ,\quad \text{%
and\quad }\partial _{t}I_{j}^{h}(t)=i\left[ I_{j}^{h}(t),h_{j}(t)\right] 
\text{.}  \label{LRin}
\end{equation}%
The solutions $\phi _{j}(t)$, $\psi _{j}(t)$ to the respective TDSEs are
related by a phase factor $\left\vert \phi _{j}(t)\right\rangle =e^{i\alpha
_{j}(t)}\left\vert \check{\phi}_{j}(t)\right\rangle $, $\left\vert \psi
_{j}(t)\right\rangle =e^{i\alpha _{j}(t)}\left\vert \check{\psi}%
_{j}(t)\right\rangle $ to the eigenstates of the invariants 
\begin{equation}
I_{j}^{h}(t)\left\vert \check{\phi}_{j}(t)\right\rangle =\Lambda
_{j}\left\vert \check{\phi}_{j}(t)\right\rangle
,~~~~~~~I_{j}^{H}(t)\left\vert \check{\psi}_{j}(t)\right\rangle =\Lambda
_{j}\left\vert \check{\psi}_{j}(t)\right\rangle ,~~~~~~~\text{with }\dot{%
\Lambda}_{j}=0.  \label{LR1}
\end{equation}%
Subsequently the phase factors can be computed from 
\begin{equation}
\dot{\alpha}_{j}=\left\langle \check{\phi}_{j}(t)\right\vert i\partial
_{t}-h_{j}(t)\left\vert \check{\phi}_{j}(t)\right\rangle =\left\langle 
\check{\psi}_{j}(t)\right\vert \eta _{j}^{\dagger }(t)\eta _{j}(t)\left[
i\partial _{t}-H_{j}(t)\right] \left\vert \check{\psi}_{j}(t)\right\rangle .
\end{equation}%
As has been shown \cite{maamache2017pseudo,AndTom4,AndTom5}, the two
invariants for the Hermitian and non-Hermitian system obeying the TDDE are
related to each other by a similarity transformation 
\begin{equation}
I_{j}^{h}=\eta _{j}I_{j}^{H}\eta _{j}^{-1}\text{.}  \label{simhH}
\end{equation}%
Here we show that the invariants $I_{0}^{H}$, $I_{1}^{H}$ and $I_{0}^{h}$, $%
I_{1}^{h}$ are related by the intertwining operators $L$ in (\ref%
{DarbouxGeneral}) and $\ell $ in (\ref{ll}), respectively. We have 
\begin{equation}
LI_{0}^{H}=I_{1}^{H}L,\qquad \text{and}\qquad \ell I_{0}^{h}=I_{1}^{h}\ell .
\label{II}
\end{equation}%
This is seen from computing%
\begin{equation}
i\partial _{t}\left( LI_{0}^{H}-I_{1}^{H}L\right) =H_{1}\left(
LI_{0}^{H}-I_{1}^{H}L\right) -\left( LI_{0}^{H}-I_{1}^{H}L\right) H_{0},
\label{LHH}
\end{equation}%
where we used (\ref{IH}) and (\ref{LRin}) to replace time-derivatives of $L$
and $I_{0}^{H}$, respectively. Comparing (\ref{LHH}) with (\ref{IH}) in the
form $i\partial _{t}L=H_{1}L-LH_{0}$, we conclude that $%
L=LI_{0}^{H}-I_{1}^{H}L$ or $LI_{0}^{H}=I_{1}^{H}L$. The second relation in (%
\ref{II}) follows from the first when using (\ref{DarbouxGeneral}) and (\ref%
{simhH}). Thus schematically the invariants are related in the same manner
as depicted for the Hamiltonians in (\ref{DH}) with the difference that the
TDDE is replaced by the simpler adjoint action of the Dyson map. Given the
above relations we have no obvious consecutive orderings of how to compute
the quantities involved. For convenience we provide a summary of the above
in the following diagram to illustrate schematically how different
quantities are related to each other:

\FIGURE{ 
\thispagestyle{empty} \setlength{\unitlength}{1.0cm} 
\begin{picture}(14.48,9.0)(-2.2,6.5)
\thicklines
\put(-0.6,12.0){\LARGE{$H_0  \quad \longleftrightarrow \quad h_0 \quad \longleftrightarrow \quad h_1 \quad \longleftrightarrow \quad H_1$}}
\put(-0.6,10.0){\LARGE{$I_0^H  \quad \longleftrightarrow \quad I_0^h \quad \longleftrightarrow \quad I_1^h \quad \longleftrightarrow \quad I_1^H$}}

\put(-0.6,8.0){\LARGE{$\check{\psi}_0 \quad \longleftrightarrow \quad \check{\phi}_0 \quad \longleftrightarrow \quad \check{\phi}_1 \quad \longleftrightarrow \quad \check{\psi}_1$}}
\put(-0.6,14.0){\LARGE{${\psi}_0 \quad \longleftrightarrow \quad {\phi}_0 \quad \longleftrightarrow \quad {\phi}_1 \quad \longleftrightarrow \quad {\psi}_1$}}

\put(-0.3,11.8){\vector(0,-1){1.2}}	
\put(3.,11.8){\vector(0,-1){1.2}}
\put(6.1,11.8){\vector(0,-1){1.2}}
\put(9.3,11.8){\vector(0,-1){1.2}}

\put(-0.3,9.7){\vector(0,-1){1.1}}	
\put(3.,9.7){\vector(0,-1){1.1}}
\put(6.1,9.7){\vector(0,-1){1.1}}
\put(9.3,9.7){\vector(0,-1){1.1}}

\put(-0.3,12.5){\vector(0,1){1.1}}	
\put(3.,12.5){\vector(0,1){1.1}}
\put(6.1,12.5){\vector(0,1){1.1}}
\put(9.3,12.5){\vector(0,1){1.1}}

\put(1.2,14.6){\LARGE{$\eta_0 $}}
\put(1.2,12.6){\LARGE{$\eta_0 $}}
\put(1.2,9.6){\LARGE{$\eta_0 $}}
\put(1.2,7.6){\LARGE{$\eta_0 $}}

\put(4.5,14.4){\LARGE{$\ell $}}
\put(4.5,12.4){\LARGE{$\ell $}}
\put(4.5,9.5){\LARGE{$\ell $}}
\put(4.5,7.5){\LARGE{$\ell $}}

\put(7.6,14.6){\LARGE{$\eta_1 $}}
\put(7.6,12.6){\LARGE{$\eta_1 $}}
\put(7.6,9.6){\LARGE{$\eta_1 $}}
\put(7.6,7.6){\LARGE{$\eta_1 $}}

\thicklines
\put(0.14,11.67){\vector(-3,2){0.2}}
\put(0.14,10.72){\vector(-3,-2){0.2}}
\put(9.0,11.68){\vector(3,2){0.2}}
\put(9.0,10.71){\vector(3,-2){0.2}}

\put(-0.9,13.8){\vector(2,3){0.2}}

\put(9.91,13.8){\vector(-2,3){0.2}}

\qbezier(0.1, 11.7)(4.8, 9.9)(9.0,11.7)
\qbezier(0.1, 10.7)(4.8, 12.5)(9.0,10.7)
\put(4.5,11.0){\LARGE{$L$}}

\qbezier(9.8, 14.0)(11.1, 11.0)(9.8,8.0)
\put(10.5,11.0){\LARGE{$\alpha_1 $}}

\qbezier(-0.8, 14.0)(-2.1, 11.0)(-0.8,8.0)
\put(-2.2,11.0){\LARGE{$\alpha_0 $}}

\end{picture}

\caption{ Schematic representation
of Dyson maps $\eta _{0}$,$\eta _{1}$ and intertwining operators $\ell $,$L$
relating quadruples of Hamiltonians $h_{0}$,$h_{1}$,$H_{0}$,$H_{1}$ and
invariants $I_{0}^{h}$,$I_{1}^{h}$,$I_{0}^{H}$,$I_{1}^{H}$ together with
their respective eigenstates $\phi _{0}$,$\phi _{1}$,$\psi _{0}$,$\psi _{1}$
and $\check{\phi}_{0}$,$\check{\phi}_{1}$,$\check{\psi}_{0}$,$\check{\psi}_{1}$ that are related by phases $\alpha _{0}$,$\alpha _{1}$.}
        \label{fig0}}

\section{Solvable potentials from the complex Gordon-Volkov Hamiltonian}

We will now discuss how the various elements in figure 1 can be computed.
Evidently the scheme allows to start from different quantities and compute
the remaining ones by following different indicated pathes, that is we may
solve intertwining relations and TDDE in different orders for different
quantities. As we are addressing here mainly the question of how to make
sense of non-Hermitian systems, we always take a non-Hermitian Hamiltonian $%
H_{0}$ as our initial starting point and given quantity. Subsequently we
solve the TDDE (\ref{Dysoneq}) for $h_{0}$,$\eta _{0}$ and thereafter close
the commutative diagrams in different ways.

We consider a complex version of the Gordon-Volkov Hamiltonian \cite{GV1,GV2}
\begin{equation}
H_{0}=H_{GV}=p^{2}+iE\left( t\right) x,
\end{equation}%
in which $iE\left( t\right) \in i\mathbb{R}$ may be viewed as a complex
electric field. In the real setting $H_{GV}$ is a Stark Hamiltonian with
vanishing potential term around which a perturbation theory can be build in
the strong field regime, see e.g. \cite{Rev1}. Such type of potentials are
also of physical interest in the study of plasmonic Airy beams in linear
optical potentials \cite{plasmonic}. Even though the Hamiltonian $H_{GV}$ is
non-Hermitian, it belongs to the interesting class of $\mathcal{PT}$%
-symmetric Hamiltonians, i.e. it remains invariant under the antilinear
transformation $\mathcal{PT}:$ $x\rightarrow -x$, $p\rightarrow p$, $%
i\rightarrow -i$.

In order to solve the TDDE (\ref{Dysoneq}) involving $H_{0}$ we make the
Ansatz 
\begin{equation}
\eta _{0}=e^{\alpha \left( t\right) x}e^{\beta \left( t\right) p},
\label{e0}
\end{equation}%
with $\alpha \left( t\right) $, $\beta \left( t\right) $ being some
time-dependent real functions. The adjoint action of $\eta _{0}$ on $x$, $p$
and the time-dependent term of Maurer-Cartan form are easily computed to 
\begin{equation}
\eta _{0}x\eta _{0}^{-1}=x-i\beta ,\quad \eta _{0}p\eta _{0}^{-1}=p+i\alpha
,\quad i\dot{\eta}_{0}\eta _{0}^{-1}=i\dot{\alpha}x+i\dot{\beta}\left(
p+i\alpha \right) .
\end{equation}%
We use now frequently overdots as an abbreviation for partial derivatives
with respect to time. Therefore the right hand side of the TDDE (\ref%
{Dysoneq}) yields 
\begin{equation}
h_{0}=h_{GV}=p^{2}+ip\left( 2\alpha +\dot{\beta}\right) -\alpha
^{2}+ix\left( E+\dot{\alpha}\right) +E\beta -\dot{\beta}\alpha .
\end{equation}%
Thus, for $h_{0}$ to be Hermitian we have to impose the reality constraints 
\begin{equation}
\dot{\alpha}=-E,\quad \dot{\beta}=-2\alpha ,
\end{equation}%
so that $h_{0}$ becomes a free particle Hamiltonian with an added real
time-dependent field 
\begin{equation}
h_{0}=h_{GV}=p^{2}+\alpha ^{2}+E\beta =p^{2}+\left[ \dint\nolimits^{t}E%
\left( s\right) ds\right] ^{2}+2E\left( t\right)
\dint\nolimits^{t}\dint\nolimits^{s}E\left( w\right) dwds.
\label{freeparticle}
\end{equation}%
There are numerous solutions to the TDSE $i\partial _{t}\phi _{0}=h_{GV}\phi
_{0}$, with each of them producing different types of partner potentials $%
v_{1}$ and hierarchies. We will now discuss various ways to construct the
next level in the hierarchy by using different types of solutions.

\subsection{Solvable time-dependent hyperbolic potentials, two separate
intertwinings}

We start by considering the scenario as depicted in the commutative diagram (%
\ref{dia}). Thus we start with a solution to the TDDE in form of $h_{0}$, $%
H_{0}$, $\eta _{0}$ as given above and carry out the intertwining relations
separately using the intertwining operators $\mathcal{L}\left[ u\right] $
and $\mathcal{L}\left[ U\right] $ in the construction of $h_{1}$ and $H_{1}$%
, respectively. According to (\ref{sc}), in this case the expression for the
second Dyson map is dictated by the closure of the diagram to be $\eta
_{1}=\eta _{0}$. We construct our intertwining operator from the simplest
solutions to the TDSE for $h_{0}=h_{GV}$%
\begin{equation}
\phi _{0,m}\left( x,t\right) =\cosh (mx)e^{-\alpha x+im^{2}t-i\int^{t}\alpha
^{2}+E\beta ds}  \label{fm}
\end{equation}%
with continuous parameter $m$. A second linearly independent solution $%
\tilde{\phi}_{0,m}$ is obtained by replacing the $\cosh $ in (\ref{fm}) by $%
\sinh $. Taking $\phi _{0,m}$ as our seed function we compute%
\begin{eqnarray}
\ell &=&\mathcal{L}\left[ \phi _{0,m}\right] =\ell _{1}\left( t\right) \left[
\partial _{x}-m\tanh (mx)\right] \\
h_{1} &=&p^{2}-2m^{2}\func{sech}^{2}(mx)+\alpha ^{2}+E\beta +i\frac{\left(
\ell _{1}\right) _{t}}{\ell _{1}} \\
\phi _{1,m,m^{\prime }} &=&\ell \lbrack \phi _{0,m}](\phi _{0,m^{\prime }})
\\
&=&\ell _{1}\left( t\right) \left[ m^{\prime }\sinh (m^{\prime }x)-m\cosh
(m^{\prime }x)\tanh (mx)\right] e^{im^{2}t-i\int^{t}\alpha ^{2}+E\beta ds}
\end{eqnarray}%
Evidently $\ell _{1}(t)$ must be constant for $h_{1}$ to be Hermitian, so
for convenience we set $\ell _{1}(t)=1$. Since $\eta _{0}$ is of the form
that solves the second equation in (\ref{sc}), we can also directly solve
the intertwining relation (\ref{IH}) for $H_{0}$ and $H_{1}$ using an
intertwining operator build from a solution for the TDSE of $H_{0}$, i.e. $%
\mathcal{L}\left[ U\right] =\mathcal{L}\left[ \eta _{0}^{-1}\phi _{0,m}%
\right] $. We obtain%
\begin{eqnarray}
H_{1} &=&p^{2}-2m^{2}\func{sech}^{2}\left[ m(x+i\beta )\right] +iE\left(
t\right) x, \\
\psi _{1,m,\tilde{m}} &=&e^{-\alpha (x+i\beta )}\phi _{1,m,\tilde{m}%
}(x+i\beta ).
\end{eqnarray}%
We verify that the TDDE for $h_{1}$ and $H_{1}$ is solved by $\eta _{1}=\eta
_{0}\,$, which is enforced by the closure of the diagram (\ref{dia}) and the
first relation in (\ref{sc}).

We can extend our analysis to the Darboux-Crum transformation and compute
the two hierarchies of solvable time-dependent hyperbolic Hamiltonians $%
H_{0} $,$H_{1}$,$H_{2}$,$\ldots $ and $h_{0}$,$h_{1}$,$h_{2}$,$\ldots $
directly from the expressions (\ref{DCa})-(\ref{H4}). For instance, we
calculate%
\begin{equation}
H_{2}=p^{2}+\frac{2(m^{2}-\tilde{m}^{2})\left[ \tilde{m}^{2}\cosh (m\hat{x}%
)-m^{2}\cosh (\tilde{m}\hat{x})\right] }{\left[ m\cosh (\tilde{m}\hat{x}%
)\sinh (m\hat{x})-\tilde{m}\cosh (m\hat{x})\sinh (\tilde{m}\hat{x})\right]
^{2}}+iE\left( t\right) x
\end{equation}%
with $\hat{x}=x+i\beta $. The solutions to the corresponding TDSE are
directly computable from the generic formula (\ref{H4}).

\subsection{Solvable time-dependent rational potentials, intertwining and
TDDE}

Next we start again with a solution to the TDDE in form of $h_{0}$, $H_{0}$, 
$\eta _{0}$, carry out the intertwining to construct $h_{1}$ and
subsequently solve the TDDE for $H_{1}$, $\eta _{1}$ with given $h_{1}$ as
depicted in the commutative diagram 
\begin{equation}
\begin{array}{ccccc}
& H_{0} & ~~\underrightarrow{\eta _{0}}~~ & h_{0} &  \\ 
? & \downarrow &  & \downarrow & \mathcal{L}\left[ u\right] \\ 
& H_{1} & \underleftarrow{\eta _{1}} & h_{1} & 
\end{array}
\label{CD1}
\end{equation}%
In this case the expression for the intertwining operator between $H_{0}$
and $H_{1}$ is dictated by the closure of the diagram to be $\eta _{1}^{-1}%
\mathcal{L}\left[ u\right] \eta _{0}\neq \mathcal{L}\left[ U\right] $. We
discuss this for a more physical solution as in the previous section that
can be found for instance in \cite{miller1977symmetry} for the free
particle, which we modify by an additional phase 
\begin{equation}
\phi _{n}^{(0)}\left( x,t\right) =\frac{1}{\left( t^{2}+1\right) ^{1/4}}%
H_{n}(iz)\exp \left[ \left( 1+it\right) z^{2}+i\kappa _{n}(t)\right]
\end{equation}%
where $z:=x/\sqrt{2+2t^{2}}$ and $\kappa _{n}(t)=\left( n+\frac{1}{2}\right)
\arctan t-\int^{t}\alpha (s)^{2}+E(s)\beta (s)ds$. There exists a more
general solution in terms of parabolic cylinder functions with a continuous
parameter, but we consider here the specialized version that only involves
Hermite polynomials $H_{n}(x)$ as this leads to more interesting potentials
of rational type. Using $\phi _{n}^{(0)}$ allows us to compute the
corresponding intertwining operators $\ell _{n}^{(1)}$ and partner
potentials $v_{n}^{(1)}$. Evaluating the formulae in (\ref{v1}) we obtain 
\begin{eqnarray}
\ell _{n}^{(1)} &=&\ell _{1}\left( t\right) \left[ -\frac{i}{2}\left( \frac{x%
}{i+t}+\frac{2n\sqrt{2}H_{n-1}(iz)}{\sqrt{1+t^{2}}H_{n}(iz)}\right)
+\partial _{x}\right] , \\
v_{n}^{(1)} &=&\frac{4n}{1+t^{2}}\left[ \frac{%
(n-1)H_{n-2}(iz)H_{n}(iz)-nH_{n-1}^{2}(iz)}{H_{n}^{2}(iz)}\right] +\alpha
^{2}+E\beta -\frac{1+it}{1+t^{2}}+i\frac{\left( \ell _{1}\right) _{t}}{\ell
_{1}}.  \notag
\end{eqnarray}%
Since the combination of Hermite polynomials in $v_{n}^{(1)}$ is always
real, we notice that $\func{Im}[v_{1}^{(n)}]$ is only a function of $t$ and
can be eliminated by a suitable choice of $\ell _{1}$. The choice (\ref{h1})
yields $\ell _{1}=\sqrt{1+t^{2}}$ for all $n$ and the rational potentials in 
$x$ and $t$ 
\begin{eqnarray}
v_{0}^{(1)} &=&\alpha ^{2}+E\beta -\frac{1}{1+t^{2}}%
,~~v_{1}^{(1)}=v_{0}^{(1)}+\frac{2}{x^{2}},~~v_{2}^{(1)}=v_{0}^{(1)}-\frac{%
4\left( 1+t^{2}-x^{2}\right) }{\left( 1+t^{2}+x^{2}\right) ^{2}}  \notag \\
v_{3}^{(1)} &=&v_{0}^{(1)}+\frac{6\left[ 3\left( 1+t^{2}\right) ^{2}+x^{4}%
\right] }{x^{2}\left( 3+3t^{2}+x^{2}\right) ^{2}},~~ \\
v_{4}^{(1)} &=&v_{0}^{(1)}+\frac{8\left[ 3\left( 1+t^{2}\right)
x^{4}+9\left( 1+t^{2}\right) ^{2}x^{2}-9\left( 1+t^{2}\right) ^{3}+x^{6}%
\right] }{\left[ 6\left( 1+t^{2}\right) x^{2}+3\left( 1+t^{2}\right)
^{2}+x^{4}\right] ^{2}},\ldots  \notag
\end{eqnarray}%
\FIGURE{\epsfig{file=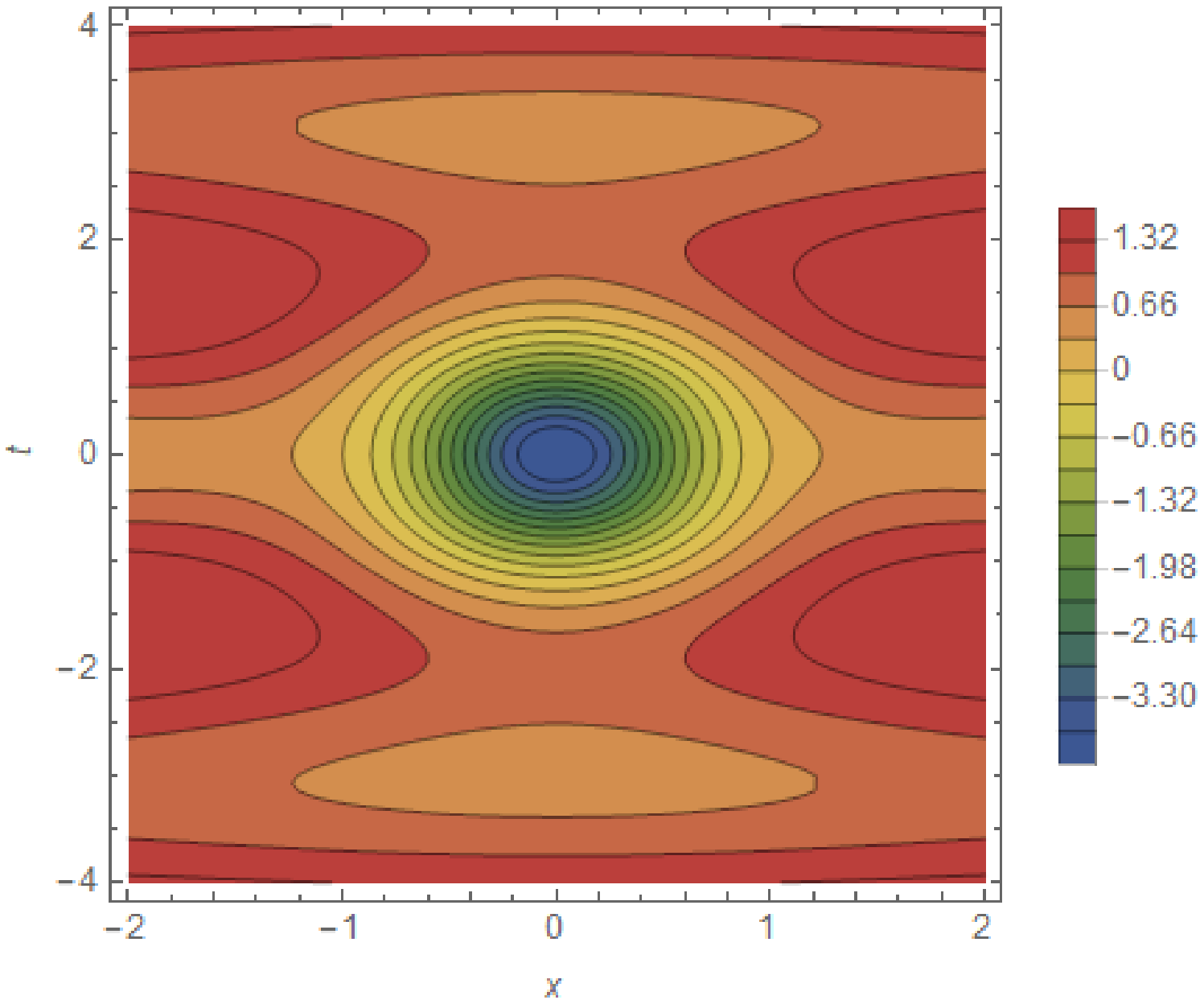,width=5.1cm}\epsfig{file=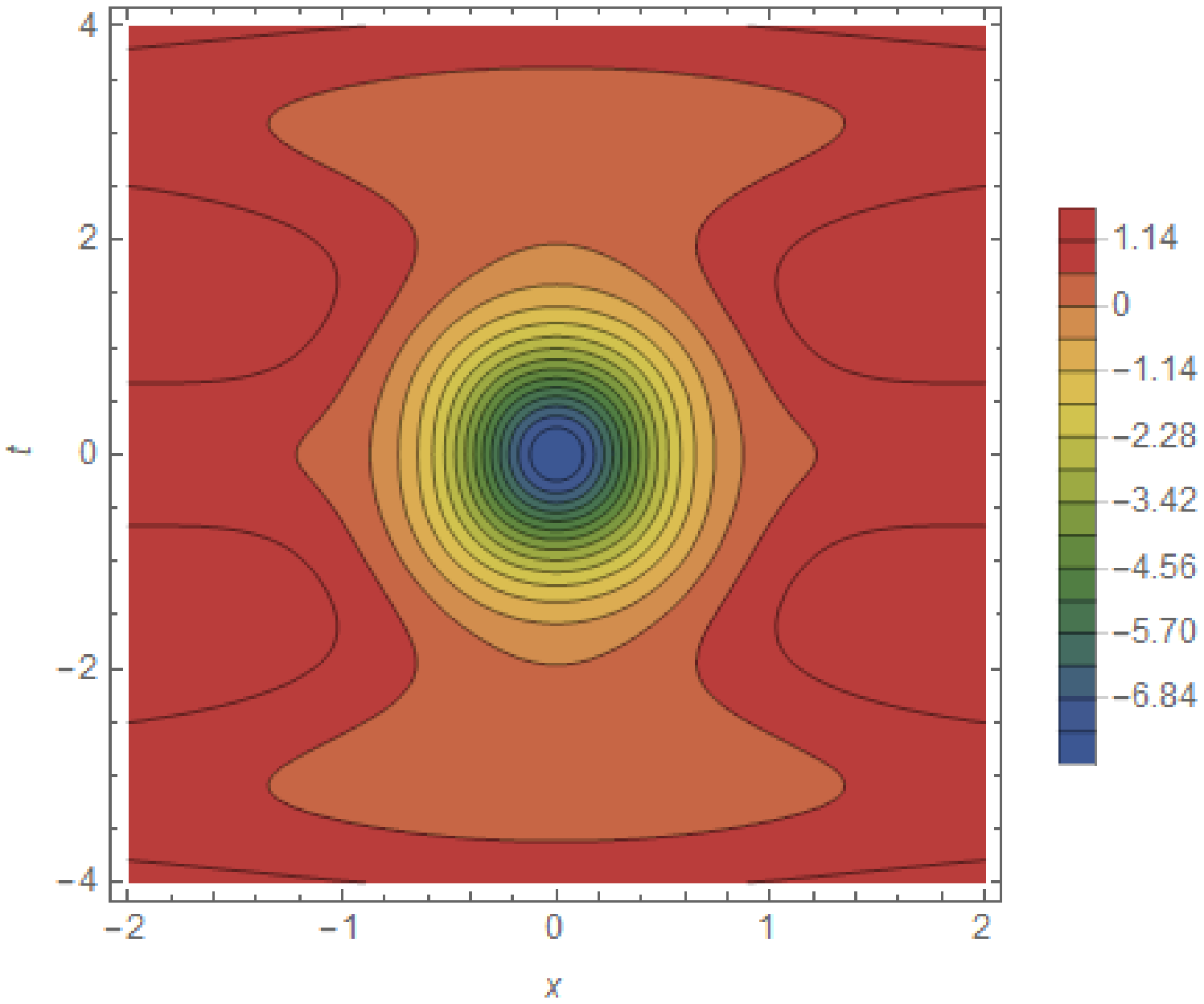,width=5.1cm}\epsfig{file=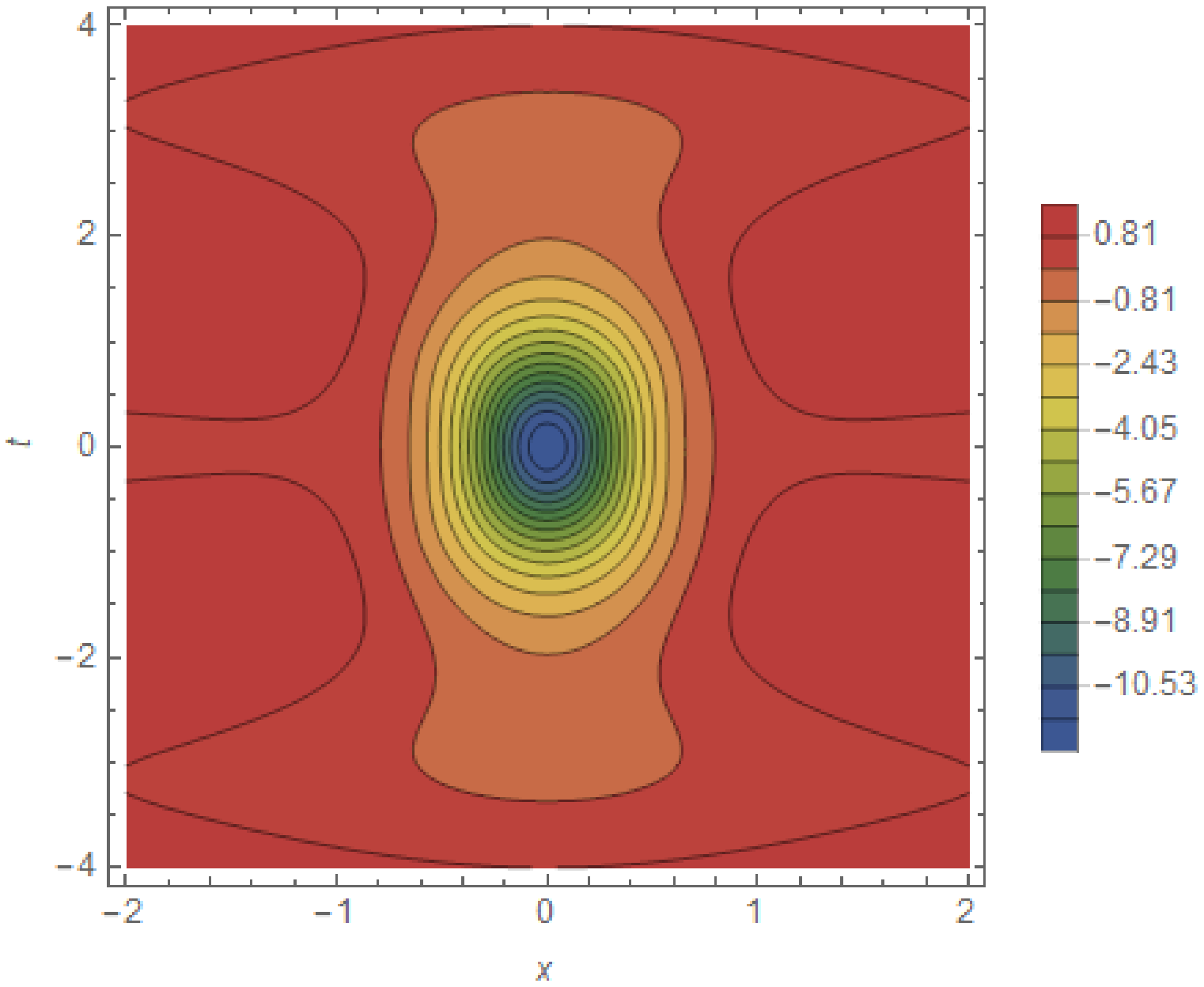,width=5.1cm} 
\caption{ Time-dependent rational potentials $v_{2}^{(1)}(x,t)$, $v_{4}^{(1)}(x,t)$ and $v_{6}^{(1)}(x,t)$ with $E(t)=\sin t$.}
        \label{fig1}}

We observe that all potentials $v_{n}^{(1)}$ with $n$ odd are singular at $%
x=0$, whereas those with $n$ even are regular for all values of $x$ and $t$.
We depict some of these finite potentials in figure \ref{fig1}, noting that
they possess well defined minima and finite asymptotic behaviour. The
nontrivial solutions (\ref{second}) to the TDSE for the Hamiltonians
involving $v_{n}^{(1)}$ are%
\begin{eqnarray}
\phi _{n,m}^{(1)} &=&\mathcal{L}\left[ \phi _{n}^{(0)}\right] \left( \phi
_{m}^{(0)}\right) ,\qquad n\neq m \\
&=&i\sqrt{2}\frac{mH_{m-1}(iz)H_{n}(iz)-nH_{m}(iz)H_{n-1}(iz)}{\left(
1+t^{2}\right) ^{1/4}H_{n}(iz)}e^{i\kappa _{m}(t)}
\end{eqnarray}%
and the nontrivial solutions obtained from (\ref{ntsol}) are 
\begin{eqnarray}
\hat{\phi}_{0}^{(1)} &=&e^{-z^{2}}\phi _{0}^{(0)},\quad \tilde{\phi}%
_{0}^{(1)}=\sqrt{2}F(z)\phi _{0}^{(0)},  \notag \\
\hat{\phi}_{1}^{(1)} &=&\frac{e^{-z^{2}}}{4z^{2}}\phi _{1}^{(0)},\quad 
\tilde{\phi}_{0}^{(1)}=\left[ x\sqrt{1+t^{2}}-\sqrt{2}(1+t^{2})F(z)\right]
\phi _{1}^{(0)}, \\
\hat{\phi}_{2}^{(1)} &=&\frac{(1+t^{2})^{2}e^{-z^{2}}}{4(1+t^{2}+x^{2})^{2}}%
\phi _{2}^{(0)},\quad \tilde{\phi}_{2}^{(1)}=\frac{\left[ x(x^{2}-t^{2}-1)%
\sqrt{1+t^{2}}-2\sqrt{2}(1+t^{2})^{2}F(z)\right] }{(1+t^{2}+x^{2})^{2}}\phi
_{2}^{(0)},  \notag
\end{eqnarray}%
where $F(z)$ denotes the Dawson integral $F(z):=\exp
(-z^{2})\dint\nolimits_{0}^{z}\exp (s^{2})ds$.

Finally we compute the non-Hermitian counterpart $H_{1}$ from the TDDE (\ref%
{Dysoneq}). Taking now $\eta _{1}$ to be of the same form as $\eta _{0}$ but
different time-dependent parameters we make the Ansatz 
\begin{equation}
\eta _{1}=e^{\gamma \left( t\right) x}e^{\delta \left( t\right) p}
\label{eta1}
\end{equation}%
and compute 
\begin{equation}
H_{1}(x,p,t)=h_{1}(x+i\delta ,p-i\gamma ,t)-i\dot{\gamma}x-i\dot{\delta}p+%
\dot{\gamma}\delta .  \label{Ham1}
\end{equation}%
Thus we obtain%
\begin{eqnarray}
H_{1,0} &=&p^{2}-2i\gamma p-\gamma ^{2}+\alpha ^{2}+E\beta -\frac{1}{1+t^{2}}%
-i\dot{\gamma}x+\dot{\gamma}\delta -i\dot{\delta}p \\
H_{1,1} &=&H_{1,0}+\frac{2}{(x+i\delta )^{2}},\quad H_{1,2}=H_{1,0}-\frac{4%
\left[ 1+t^{2}-(x+i\delta )^{2}\right] }{\left[ 1+t^{2}+(x+i\delta )^{2}%
\right] ^{2}}, \\
H_{1,3} &=&H_{1,0}+\frac{6\left[ 3\left( 1+t^{2}\right) ^{2}+(x+i\delta )^{4}%
\right] }{(x+i\delta )^{2}\left[ 3+3t^{2}+(x+i\delta )^{2}\right] ^{2}}%
,\ldots
\end{eqnarray}%
By setting $\dot{\delta}=-2\gamma $ we may remove the linear term in $p$ and
convert the Hamiltonian into a potential one. We notice that the
singularities for $v_{1,n}$ with $n$ odd have been regularized in the
non-Hermitian setting for $\delta \neq 0$. The remaining factors lead to
further restrictions for $\delta $ when demanding regularity for the $%
H_{1,n} $. In this case we require in addition $|\delta |<1$ for $n=2$, $%
|\delta |<\sqrt{3}$ for $n=3$, $|\delta |>\sqrt{3+\sqrt{6}}$ for $n=4$%
,\ldots\ 

We verify that according to the commutative diagram (\ref{CD1}) the
intertwining operator relating $H_{0}$ and $H_{1}$ in (\ref{IH}) is indeed $%
L=\eta _{1}^{-1}\mathcal{L}\left[ u\right] \eta _{0}$. From this we can now
also compute the nontrivial solutions (\ref{ntP}) to the TDSE%
\begin{equation}
\psi _{1}(x,t)=e^{-\gamma x-i\gamma \delta }\hat{\phi}_{1}(x+i\delta
),\qquad \text{and\qquad }\tilde{\psi}_{1}(x,t)=e^{-\gamma x-i\gamma \delta }%
\tilde{\phi}_{1}(x+i\delta ).
\end{equation}%
Hence all of these systems are exactly solvable and the diagram (\ref{CD1})
does indeed close. The two hierarchies of solvable time-dependent rational
Hamiltonians are then directly computed from the expressions (\ref{DCa})-(%
\ref{H4}).

\subsection{Solvable time-dependent Airy function potentials, two
intertwinings}

Finally we start again with a solution to the TDDE for $h_{0}$, $H_{0}$, $%
\eta _{0}$ and carry out the intertwining relations separately constructing $%
h_{1}$, $H_{1}$, but unlike as in section 4.1 we use the intertwining
operator $L=\eta _{1}^{-1}\mathcal{L}\left[ u\right] \eta _{0}$ involving an
arbitrary operator $\eta _{1}$, 
\begin{equation}
\begin{array}{ccccc}
& H_{0} & ~~\underrightarrow{\eta _{0}}~~ & h_{0} &  \\ 
L=\eta _{1}^{-1}\mathcal{L}\left[ u\right] \eta _{0} & \downarrow &  & 
\downarrow & \mathcal{L}\left[ u\right] \\ 
& H_{1} & \underrightarrow{?} & h_{1} & 
\end{array}
\label{CD2}
\end{equation}%
which, by the closure of the diagram, must be the Dyson map for the system $%
1 $.

We discuss this scenario for a somewhat less well known solution to the free
particle TDSE in terms of Airy packet solutions as found forty years ago by
Berry and Balazs \cite{berry1979}, see also \cite{gori1999general} for a
different approach. The interesting feature of these wave packets is that
they continually accelerate in a shape-preserving fashion despite the fact
that no force is acting on them. Only more recently such type of waves have
been realized experimentally in various forms, e.g. \cite%
{siviloglou2007o,siviloglou2007a,baumgartl2008,vettenburg2014light,patsyk2018}%
. As in the previous section we modify the standard solution by a phase so
that it solves the TDSE for $h_{GV}$ 
\begin{equation}
\phi _{0}^{X}\left( x,t\right) =\limfunc{Xi}\left( \gamma x-\gamma
^{4}t^{2}\right) \exp \left[ i\gamma ^{3}t\left( x-\frac{2\gamma ^{3}t^{2}}{3%
}\right) -i\int \alpha ^{2}+E\beta dt\right] .
\end{equation}%
Here $\limfunc{Xi}\left( z\right) $ denotes any of the two Airy functions $%
\limfunc{Ai}\left( z\right) $ or $\limfunc{Bi}\left( z\right) $ and $\gamma
\in \mathbb{C}$ is a free parameter. Using once more the relation in (\ref%
{v1}), we obtain the intertwining operators and new Hamiltonians 
\begin{eqnarray}
\ell ^{X} &=&\ell _{1}\left( t\right) \left[ -i\gamma t^{3}-\gamma \frac{%
\limfunc{Xi}^{\prime }\left( \gamma x-\gamma ^{4}t^{2}\right) }{\limfunc{Xi}%
\left( \gamma x-\gamma ^{4}t^{2}\right) }+\partial _{x}\right] , \\
h_{1}^{X} &=&2\gamma ^{3}(x-\gamma ^{3}t^{2})-2\gamma ^{2}\left[ \frac{%
\limfunc{Xi}^{\prime }\left( \gamma x-\gamma ^{4}t^{2}\right) }{\limfunc{Xi}%
\left( \gamma x-\gamma ^{4}t^{2}\right) }\right] ^{2}+\alpha ^{2}+E\beta +i%
\frac{\left( \ell _{1}\right) _{t}}{\ell _{1}},  \notag
\end{eqnarray}%
with $\limfunc{Xi}^{\prime }\left( z\right) $ denoting the derivative of the
Airy functions. Taking $\ell _{1}$ to be a constant and $\gamma \in \mathbb{R%
}$ these are indeed Hermitian Hamiltonians. We also note that $h_{1}^{X}$
becomes singular when $\gamma x-\gamma ^{4}t^{2}$ equals a zero of the Airy
functions on the negative real axis. In addition, $h_{1}^{B}$ becomes
singular when $\pi /3<\arg (\gamma x-\gamma ^{4}t^{2})<\pi /2$. The
nontrivial solutions according to (\ref{second}) are computed to%
\begin{equation}
\phi _{1}^{A/B}=\ell ^{A/B}(\phi _{0}^{B/A})=\pm \frac{\ell _{1}(t)\gamma
\exp \left[ -\frac{1}{3}i\left( 2\gamma ^{6}t^{3}-3\gamma ^{3}tx+3\int^{t}%
\left[ \alpha (s)^{2}+\beta (s)\kappa (s)\right] \,ds\right) \right] }{\pi 
\limfunc{Ai}/\limfunc{Bi}\left( \gamma x-\gamma ^{4}t^{2}\right) }.
\end{equation}%
We have constructed these solutions from the two linearly independent
solutions to the original TDSE rather than from one particular solution with
different parameters $\gamma $, i.e.%
\begin{equation}
\phi _{1}^{A/B}(\gamma _{1},\gamma _{2})=\mathcal{L}[\phi _{0}^{A/B}(\gamma
_{1})](\phi _{0}^{A/B}(\gamma _{2}))
\end{equation}%
are also solutions. Additional solutions can also be obtained in a
straightforward manner from (\ref{ntsol}).

For fixed values of time we observe in figure \ref{fig3} panel (a) the two
characteristic qualitatively different types of behaviour of the Airy wave
function, that is being oscillatory up to a certain point $x=x_{0}$ and
beyond which the density distribution becomes decaying. We observe further
that for increasing positive time, or decreasing negative time, the wave
packets accelerate. For the density wave function of the partner Hamiltonian
in panel (b) we observe this behaviour for one dominating value of $\gamma $
modulated by the other.

\FIGURE{\epsfig{file=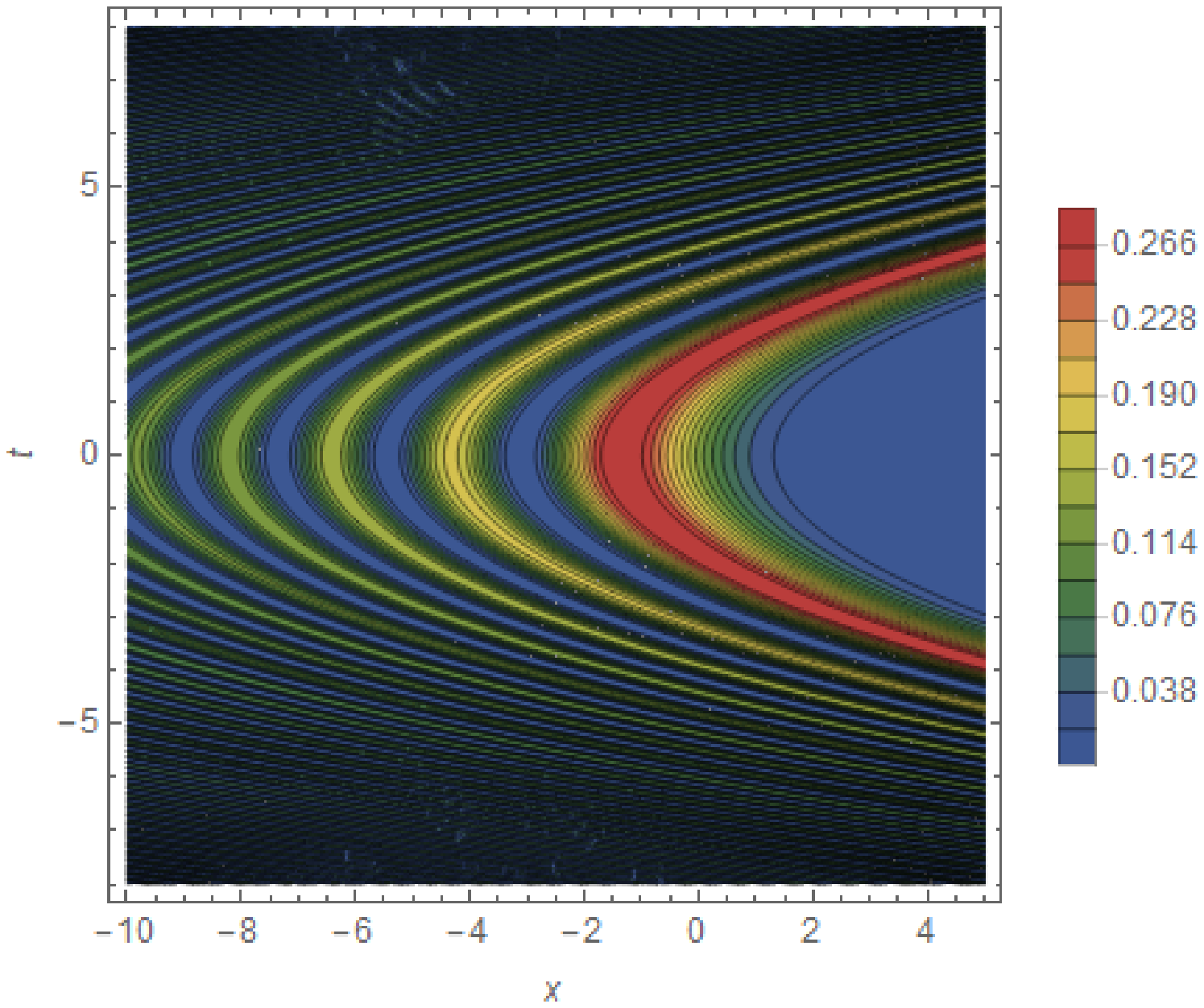,width=7.1cm} \epsfig{file=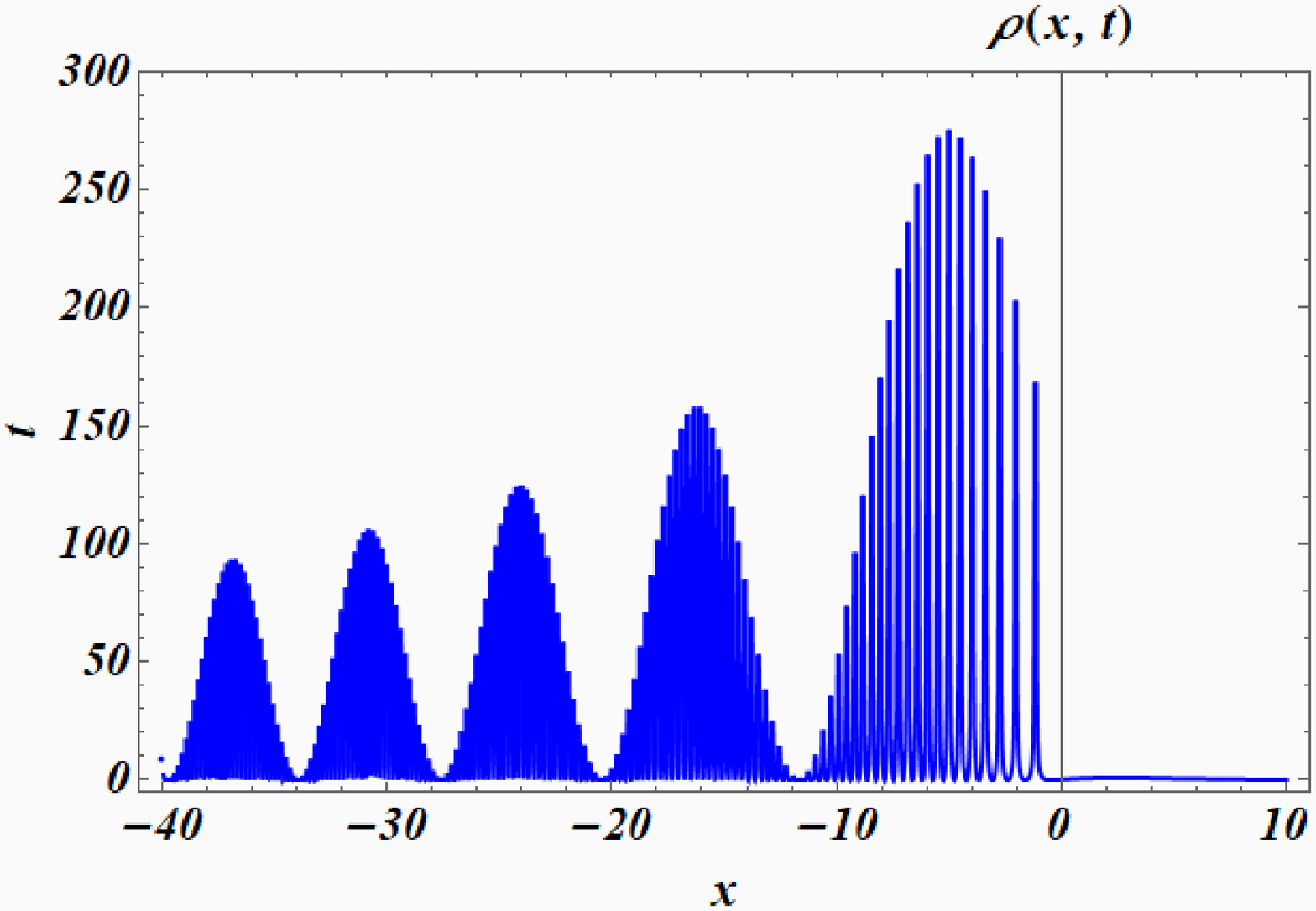,width=7.1cm}
\caption{Probability densities for Airy wavepackets for solutions of the level 1 and 2 TDSE $\rho_0 =\left\vert \phi
_{0}^{A}(\gamma =0.75)\right\vert ^{2}$ and $\rho_1 =\left\vert \phi
_{1}^{A}(t=1, \gamma _{1}=0.2,\gamma _{2}=2.0)\right\vert ^{2}$, left and right
panel, respectively.}
        \label{fig3}}

According to our commutative diagram (\ref{CD2}) we calculate next the
non-Hermitian counterpart $H_{1}^{X}$ using the intertwining operator $%
L=\eta _{1}^{-1}\mathcal{L}\left[ \phi _{0}^{X}\right] \eta _{0}$ with $\eta
_{1}$ as specified in (\ref{eta1}). We obtain 
\begin{equation}
H_{1}^{X}(x,p,t)=h_{1}^{X}(x+i\delta ,p-i\gamma ,t)-i\dot{\gamma}x-i\dot{%
\delta}p+\dot{\gamma}\delta .
\end{equation}%
We verify the closure of the diagram by noting that $H_{1}^{X}$ satisfies
indeed the TDDE with $h_{1}^{X}$, $\eta _{1}$.

The above mentioned singularities on the real axis are now regularized.

\section{Reduced Swanson model hierarchy}

Next we consider a model that is build from a slightly more involved
time-dependent Dyson map. We proceed as outlined in the commutative diagram (%
\ref{CD1}). Our simple starting point is a non-Hermitian, but $\mathcal{PT}$%
-symmetric, Hamiltonian that may be viewed as reduced version of the
well-studied Swanson model \cite{Swanson} 
\begin{equation}
H_{0}=H_{RS}=ig\left( t\right) xp.
\end{equation}%
We follow the same procedure as before and solve at first the TDDE for $\eta
_{0}$ and $h_{0}$ with given $H_{0}$. In this case the arguments in the
exponentials of the time-dependent Dyson map can no longer be linear and we
therefore make the Ansatz 
\begin{equation}
\eta _{0}=e^{\lambda \left( t\right) xp}e^{\zeta \left( t\right) p^{2}/2}.
\label{eta0}
\end{equation}%
The right hand side of the TDDE (\ref{Dysoneq}) is then computed to 
\begin{equation}
h_{0}=h_{RS}=\left[ \left( g\zeta +i\frac{\dot{\zeta}}{2}\right) \cos
(2\lambda )+\left( ig\zeta -\frac{\dot{\zeta}}{2}\right) \sin (2\lambda )%
\right] p^{2}+i(g+\dot{\lambda})xp.
\end{equation}%
Thus for $h_{0}$ to be Hermitian we have to impose 
\begin{equation}
\dot{\lambda}=-g,\quad \dot{\zeta}=-2g\zeta \tan 2\lambda ,  \label{const}
\end{equation}%
so that we obtain a free particle Hamiltonian with a time-dependent mass $%
m(t)$%
\begin{equation}
h_{0}=h_{RS}=\frac{1}{2m(t)}p^{2},\qquad \text{with }m(t)=\frac{1}{2g\zeta
\sec (2\lambda )}.  \label{freep}
\end{equation}%
Time-dependent masses have been proposed as a possible mechanism to explain
anomalous nuclear reactions which can not been explained by existing
conventional theories in nuclear physics, see e.g. \cite{masst}. The reality
constraints (\ref{const}) can be solved by 
\begin{equation}
\lambda (t)=-\int^{t}g\left( s\right) ds,\quad \text{and \quad }\zeta
(t)=c\sec \left( 2\int^{t}g\left( s\right) ds\right) ,\quad
\end{equation}%
with constant $c$. Thus the time-dependent mass $m(t)$ can be expressed
entirely in terms of the time-dependent coupling $g(t)$. An exact solution
to the TDSE for $h_{RS}$ can be found for instance in \cite{pedrosa1997exact}
when setting in there the time-dependent frequency to zero%
\begin{eqnarray}
\phi _{n}^{(0)}\left( x,t\right) &=&\frac{e^{i\alpha _{0,n}(t)}}{\pi ^{1/4}%
\sqrt{n!2^{n}\varrho (t)}}\exp \left[ m(t)\left( i\frac{\dot{\varrho}(t)}{%
\varrho (t)}-\frac{1}{m(t)\varrho ^{2}(t)}\right) \frac{x^{2}}{2}\right]
H_{n}\left[ \frac{x}{\varrho (t)}\right] ,  \label{Ped} \\
\alpha _{0,n}(t) &=&~\ -\dint\nolimits_{0}^{t}\frac{\left( n+1/2\right) }{%
m(s)\varrho ^{2}(s)}ds.  \label{al}
\end{eqnarray}%
For (\ref{Ped}) to be a solution, the auxiliary function $\varrho (t)$ needs
to obey the dissipative Ermakov-Pinney equation with vanishing linear term 
\begin{equation}
\ddot{\varrho}+\frac{\dot{m}}{m}\dot{\varrho}=\frac{1}{m^{2}\varrho ^{3}}.
\label{EP2}
\end{equation}%
We derive an explicit solution for this equation in appendix A. Evaluating
the formulae in (\ref{v1}), with $h_{0}$ and $h_{1}$ divided by $2m(t)$, we
obtain the intertwining operators and the partner Hamiltonians 
\begin{eqnarray}
\ell _{n}^{(1)} &=&\ell _{1}\left( t\right) \left[ \frac{x}{\varrho ^{2}}-%
\frac{2nH_{n-1}\left[ x/\varrho \right] }{\varrho H_{n}\left[ x/\varrho %
\right] }-ixm\frac{\dot{\varrho}}{\varrho }+\partial _{x}\right] ,
\label{ln} \\
h_{1,n} &=&h_{0}+\frac{4n}{m\varrho ^{2}}\left[ \frac{nH_{n-1}^{2}\left[
x/\varrho \right] -(n-1)H_{n-2}\left[ x/\varrho \right] H_{n}\left[
x/\varrho \right] }{H_{n}^{2}\left[ x/\varrho \right] }\right] +\frac{1}{%
m\varrho ^{2}}+i\left[ \frac{\dot{\ell}_{1}}{\ell _{1}}-\frac{\dot{\varrho}}{%
\varrho }\right] ,~~~  \notag  \label{hn}
\end{eqnarray}%
respectively. As in the previous section, the imaginary part of the
Hamiltonian only depends on time and can be made to vanish with the suitable
choice of $\ell _{1}=\varrho $. For concrete values of $n$ we obtain for
instance the time-dependent Hermitian Hamiltonians 
\begin{eqnarray}
h_{1,0} &=&\frac{p^{2}}{2m}+\frac{1}{m\varrho ^{2}},\qquad h_{1,1}=h_{1,0}+%
\frac{1}{mx^{2}},\qquad h_{1,2}=h_{1,0}+\frac{4(\varrho ^{2}+2x^{2})}{%
m(\varrho ^{2}-2x^{2})^{2}}, \\
h_{1,3} &=&h_{1,0}+\frac{3(3\varrho ^{4}+4x^{4})}{m(2x^{3}-3x\varrho
^{2})^{2}},\qquad h_{1,4}^{(4)}=h_{1,0}+\frac{8\left( 9\varrho
^{6}-12x^{4}\varrho ^{2}+18x^{2}\varrho ^{4}+8x^{6}\right) }{m\left(
3\varrho ^{4}-12x^{2}\varrho ^{2}+4x^{4}\right) ^{2}}.~~~~~~
\end{eqnarray}%
Notice that all these Hamiltonians are singular at certain values of $x$ and 
$t$ as $\varrho $ is real. Solutions to the TDSE for the Hamiltonian $%
h_{1,n} $ can be computed according to (\ref{second})%
\begin{equation}
\phi _{n,k}^{(1)}\left( x,t\right) =\ell _{n}^{(1)}(\phi _{k}^{(0)})=\frac{%
2^{3/2}}{\sqrt{k-n}}\left[ \frac{kH_{k-1}\left[ x/\varrho \right] }{H_{k}%
\left[ x/\varrho \right] }-\frac{nH_{n-1}\left[ x/\varrho \right] }{H_{n}%
\left[ x/\varrho \right] }\right] \phi _{k}^{(0)},~~\ n\neq k.
\end{equation}%
Both $\phi _{n}^{(0)}$ and $\phi _{n,k}^{(1)}$ are square integrable
functions with $L^{2}(\mathbb{R})$-norm equal to $1$. In figure \ref{fig4}
we present the computation for some typical probability densities obtained
from these functions. Notice that demanding $m(t)>0$ we need to impose some
restrictions for certain choices of $g(t)$.

\FIGURE{\epsfig{file=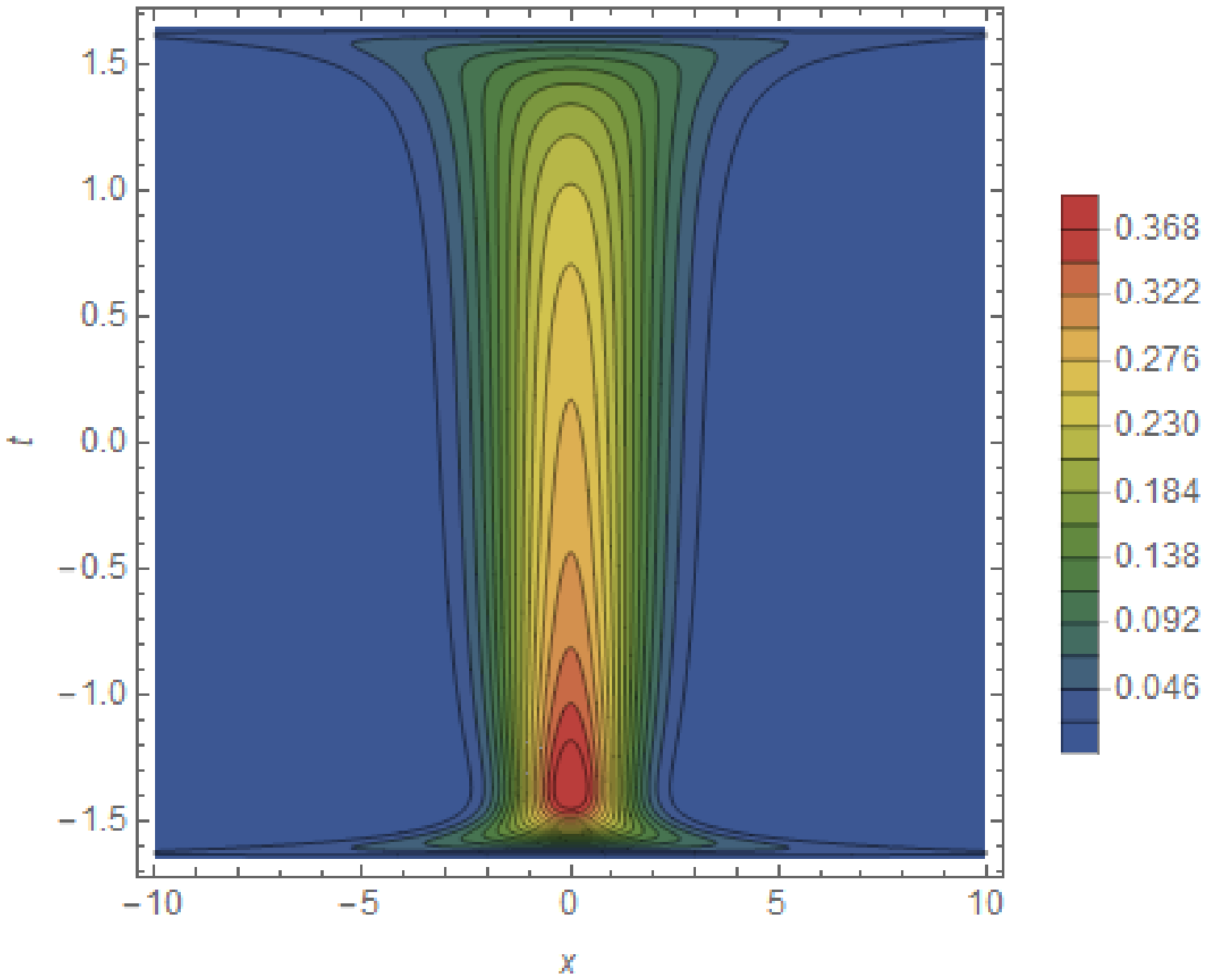,width=5.1cm}\epsfig{file=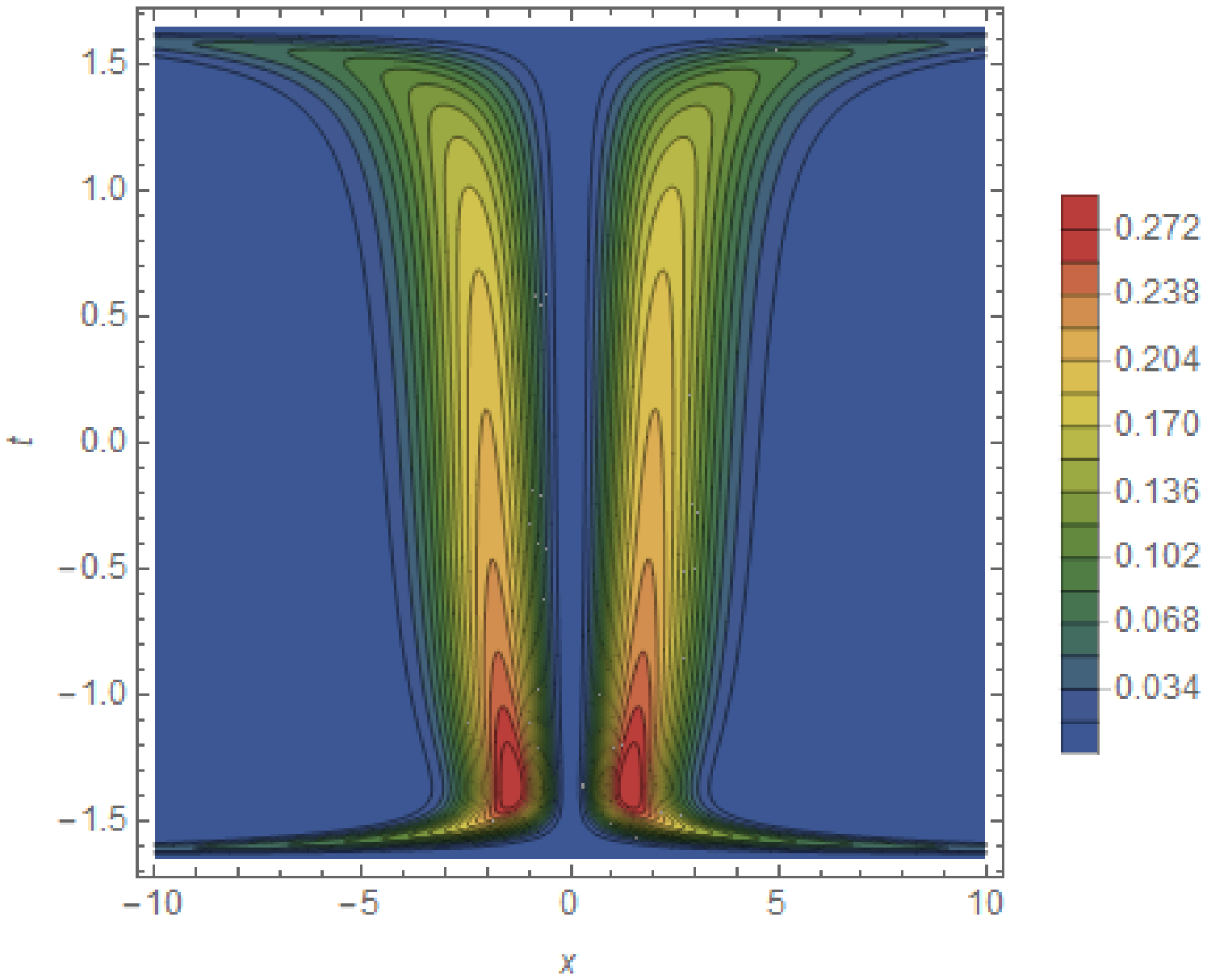,width=5.1cm}\epsfig{file=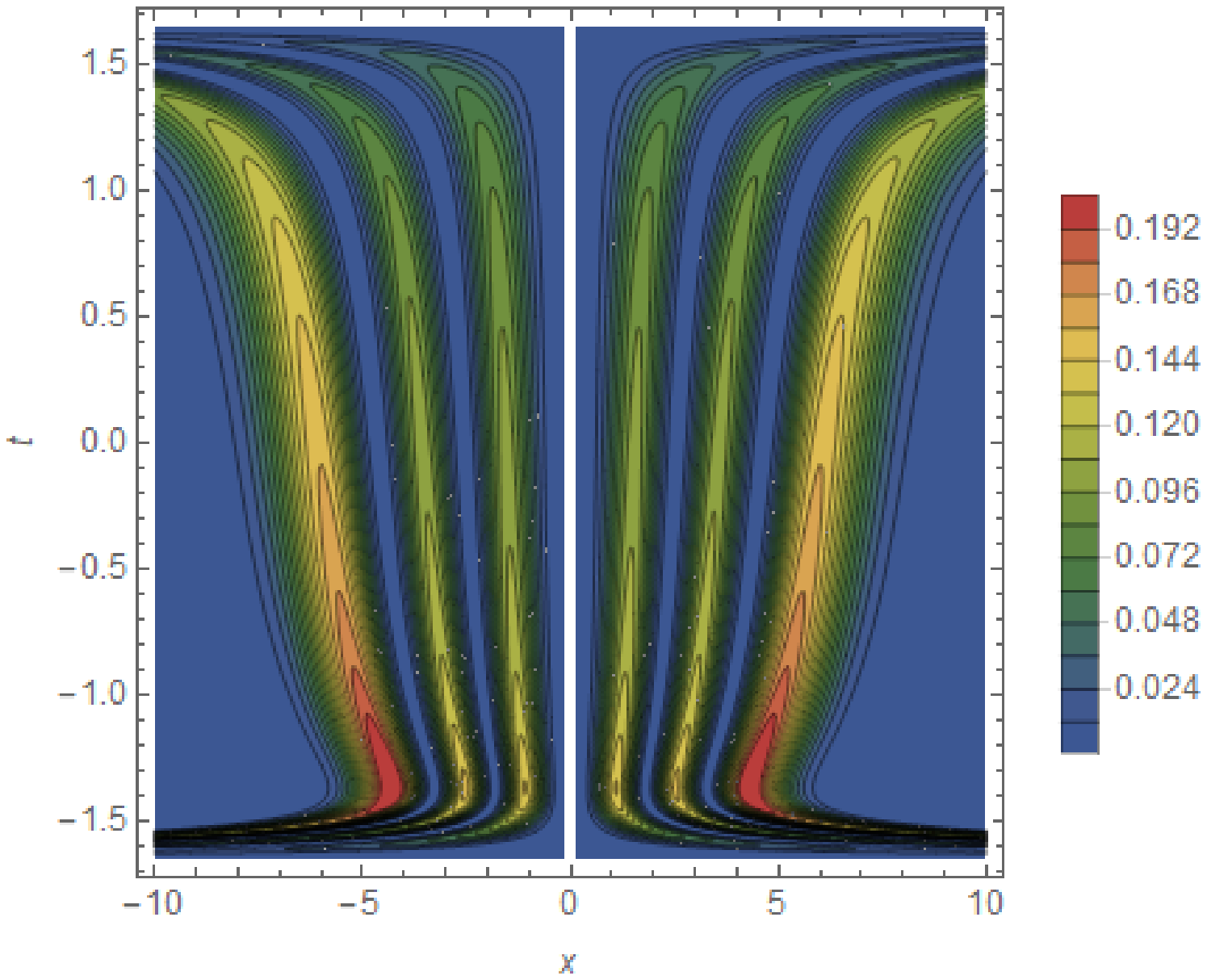,width=5.1cm} 
\caption{Probability densities $\left\vert \phi _{0}^{(0)}\right\vert ^{2}$, $%
\left\vert \phi _{1}^{(0)}\right\vert ^{2}$, $\left\vert \phi
_{1,7}^{(1)}\right\vert ^{2}$ from left to right for $g(t)=(1+t^{2})/4$, $%
m(t)=\left[ 1+\cos (t+t^{3}/3)\right] /(1+t^{2})$, $\varrho (t)=\sqrt{%
1+[C+B\tan (t/2+t^{3}/6)]^{2}}$ with $B=1/2$ and $C=1$. }
        \label{fig4}}

Next we compute the non-Hermitian counterpart $H_{1\text{ }}$with a concrete
choice for the second Dyson map. Taking $\eta _{1}$ for instance to be of
the same form as in (\ref{eta1}) the non-Hermitian Hamiltonian is formally
the same as in equation (\ref{Ham1}). In our concrete case we obtain for
instance 
\begin{equation}
H_{1,1}=\frac{p^{2}}{2m}+\frac{1}{m(x+i\delta )^{2}}-i\dot{\gamma}x+\frac{1}{%
m\varrho ^{2}}-\frac{\gamma ^{2}}{2m}+\dot{\gamma}\delta ,
\end{equation}%
where we have also imposed the constraint $\dot{\delta}=-\gamma /m$ to
eliminate a linear term in $p$, hence making the Hamiltonian a potential
one. The solutions for the TDSEs for $H_{0}$ and $H_{1,n}$ are%
\begin{equation}
\psi _{n}^{(0)}=\eta _{0}^{-1}\phi _{n}^{(0)},\quad \text{and\quad }\psi
_{n,k}^{(1)}=\eta _{1}^{-1}\phi _{n,k}^{(1)},
\end{equation}%
respectively.

\subsection{Lewis-Riesenfeld invariants}

Having solved the TDDE for $\eta _{0}$ and $\eta _{1}$ we can now also
verify the various intertwining relations for the Lewis-Riesenfeld
invariants as derived in section 3. We proceed here as depicted in the
following commutative diagram 
\begin{equation}
\begin{array}{ccccc}
& I_{0}^{h} & ~~\underrightarrow{\eta _{0}}~~ & I_{0}^{H} &  \\ 
\mathcal{L}\left[ \check{\phi}\right] & \downarrow &  & \downarrow & ? \\ 
& I_{1}^{h} & \underrightarrow{\eta _{1}} & I_{1}^{H} & 
\end{array}
\label{di2}
\end{equation}%
See also the more general schematic representation in figure \ref{fig0}. We
start with the Hermitian invariant $I_{0}^{h}$ from which we compute the
non-Hermitian invariant $I_{0}^{H}$ using the Dyson map $\eta _{0}$ as
specified in (\ref{eta0}). Subsequently we use the intertwining operator $%
\ell _{n}^{(1)}$ in (\ref{ln}) to compute the Hermitian invariants $%
I_{1,n}^{h}$ for the Hamiltonians $h_{1,n}$. The invariant $I_{1}^{H}$ is
then computed from the adjoint action of $\eta _{1}^{-1}$ as specified in (%
\ref{eta1}). Finally, the intertwining relation between the non-Hermitian
invariants $I_{0}^{H}$ and $I_{1}^{H}$ is just given by the closure of the
diagram in (\ref{di2}).

The invariant for the Hermitian Hamiltonian $h_{0}$ has been computed
previously in \cite{pedrosa1997exact}\footnote{%
We corrected a small typo in there and changed the power $1/2$ on the $%
x/\rho $-term into $2$.}%
\begin{equation}
I_{0}^{h}=A_{h}(t)p^{2}+B_{h}(t)x^{2}+C_{h}(t)\{x,p\},
\end{equation}%
where the time-dependent coefficients are%
\begin{equation}
A_{h}=\frac{\varrho ^{2}}{2},\quad B_{h}=\frac{1}{2}\left( \frac{1}{\varrho
^{2}}+m^{2}\dot{\varrho}^{2}\right) ,\quad C_{h}=-\frac{1}{2}m\varrho \dot{%
\varrho}.
\end{equation}%
It then follows from%
\begin{equation}
\left[ I_{0}^{h},h_{0}\right] =\frac{2i}{m}\left( C_{h}p^{2}+\frac{1}{2}%
B_{h}\{x,p\}\right) ,\quad \dot{A}_{h}=-\frac{2}{m}C_{h},\quad \dot{B}%
_{h}=0,\quad \dot{C}_{h}=-\frac{1}{m}B_{h},
\end{equation}%
that the defining relation (\ref{LRin}) for the invariant is satisfied by $%
I_{0}^{h}$. According to the relation (\ref{simhH}), the non-Hermitian
invariant $I_{0}^{H}$ for the non-Hermitian Hamiltonian $H_{0}$ is simply
computed by the adjoint action of $\eta _{0}^{-1}$ on $I_{0}^{h}$. Using the
expression (\ref{eta0}) we obtain%
\begin{equation}
I_{0}^{H}=\eta _{0}^{-1}I_{0}^{h}\eta
_{0}=A_{H}(t)p^{2}+B_{H}(t)x^{2}+C_{H}(t)\{x,p\},
\end{equation}%
with%
\begin{equation}
A_{H}=\frac{1}{2}e^{-2i\lambda }\rho ^{2}-\zeta ^{2}B_{H}-i\zeta m\rho \dot{%
\rho},\quad B_{H}=\frac{e^{2i\lambda }\left( 1+m^{2}\rho ^{2}\dot{\rho}%
^{2}\right) }{2\rho ^{2}},\quad C_{H}=i\zeta B_{H}-\frac{1}{2}m\rho \dot{\rho%
}.
\end{equation}%
We verify that $I_{0}^{H}$ is indeed an invariant for $H_{0}$ according to
the defining relation (\ref{LRin}), by computing%
\begin{equation}
\left[ I_{0}^{H},H_{0}\right] =2g\left( A_{H}p^{2}-B_{H}x^{2}\right) ,\quad 
\dot{A}_{H}=2igA_{H},\quad \dot{B}_{h}=-2igB_{H},\quad \dot{C}_{H}=0,
\end{equation}%
using the constraints (\ref{const}) and (\ref{EP2}).

Given the intertwining operators $\ell _{n}^{(1)}$ in (\ref{ln}) and the
invariant $I_{0}^{h}$, we can use the intertwining relation (\ref{II}) to
compute the invariants $I_{1,n}^{h}$ for the Hamiltonians $h_{1,n}$ in (\ref%
{hn}). Solving (\ref{II}) we find%
\begin{equation}
I_{1,n}^{h}=I_{0}^{h}+1+4n^{2}\frac{H_{n-1}^{2}\left[ x/\varrho \right] ^{2}%
}{H_{n}^{2}\left[ x/\varrho \right] ^{2}}-4n(n-1)\frac{H_{n-2}\left[
x/\varrho \right] }{H_{n}^{2}\left[ x/\varrho \right] }.
\end{equation}%
We verify that this expression solves (\ref{LRin}). The last invariant in
our quadruple is%
\begin{equation}
I_{1,n}^{H}(x,p)=\eta _{1}^{-1}I_{1,n}^{h}(x,p)\eta
_{1}=I_{1,n}^{h}(x+i\delta ,p-i\gamma )
\end{equation}%
Finally we may also verify the eigenvalue equations for the four invariants.
Usually this is of course the first consideration as the whole purpose of
employing Lewis-Riesenfeld invariants is to reduce the TDSE to the much
easier to solve eigenvalue equations. Here this computation is simply a
consistency check. With%
\begin{eqnarray}
\check{\phi}_{n}^{(0)} &=&e^{-i\alpha _{0,n}}\phi _{n}^{(0)},\quad \check{%
\phi}_{n,m}^{(1)}=e^{-i\alpha _{0,m}}\phi _{n,m}^{(1)},\quad \\
\check{\psi}_{n}^{(0)} &=&e^{-i\alpha _{0,n}}\psi _{n}^{(0)},\quad \check{%
\psi}_{n,m}^{(1)}=e^{-i\alpha _{0,m}}\psi _{n,m}^{(1)},
\end{eqnarray}%
and $\alpha _{0,n}$ as specified in equation (\ref{al}) we compute 
\begin{eqnarray}
I_{0}^{h}\check{\phi}_{n}^{(0)} &=&\left( n+1/2\right) \check{\phi}%
_{n}^{(0)},~~~\ \ \ I_{1,n}^{h}\check{\phi}_{n,m}^{(1)}=\left( m+1/2\right) 
\check{\phi}_{n,m}^{(1)},~~ \\
I_{0}^{H}~\check{\psi}_{n}^{(0)} &=&\left( n+1/2\right) \check{\psi}%
_{n}^{(0)},~~~\ \ \ I_{1,n}^{H}\check{\psi}_{n,m}^{(1)}=\left( m+1/2\right) 
\check{\psi}_{n,m}^{(1)}.
\end{eqnarray}%
As expected all eigenvalues are time-independent.

\section{Conclusions}

We have generalized the scheme of time-dependent Darboux transformations to
allow for the treatment of non-Hermitian Hamiltonians that are $\mathcal{PT}$%
-symmetric/quasi-Hermitian. It was essential to employ intertwining
operators different from those used in the Hermitian scheme previously
proposed. We have demonstrated that the quadruple of Hamiltonians, two
Hermitian and two non-Hermitian ones, can be constructed in alternative
ways, either by solving two TDDEs and one intertwining relation or by
solving one TDDE and two intertwining relations. For a special class of
Dyson maps it is possible to independently carry out the intertwining
relations for the Hermitian and non-Hermitian sector, which, however, forced
the seed function used in the construction of the intertwining operator to
obey certain constraints. We extended the scheme to the construction of the
entire time-dependent Darboux-Crum hierarchies. We also showed that the
scheme is consistently adaptable to construct Lewis-Riesenfeld invariants by
means of intertwining relations. Here we verified this for a concrete system
by having already solved the TDSE, however, evidently it should also be
possible to solve the eigenvalue equations for the invariants first and
subsequently construct the solutions to the TDSE. As in the Hermitian case,
our scheme allows to treat time-dependent systems directly instead of having
to solve the time-independent system first and then introducing time by
other means. The latter is not possible in the context of the Schr\"{o}%
dinger equation, unlike as in the context of nonlinear differential
equations that admit soliton solutions, where a time-dependence can be
introduced by separate arguments, such as for instance using Galilean
invariance. Naturally it will be very interesting to apply our scheme to the
construction of multi-soliton solutions.

\appendix

\section{Appendix}

We briefly explain how to solve the Ermakov-Pinney equation with dissipative
term (\ref{EP2})%
\begin{equation}
\ddot{\varrho}+\frac{\dot{m}}{m}\dot{\varrho}=\frac{1}{m^{2}\varrho ^{3}}.
\label{A1}
\end{equation}%
The solutions to the standard version of the equation \cite{Ermakov,Pinney} 
\begin{equation}
\ddot{\sigma}+\lambda (t)\sigma =\frac{1}{\sigma ^{3}}  \label{EPtime}
\end{equation}%
are well known to be of the form \cite{Pinney}%
\begin{equation}
\sigma (t)=\left( Au^{2}+Bv^{2}+2Cuv\right) ^{1/2},  \label{EPsol}
\end{equation}%
with $u(t)$ and $v(t)$ denoting the two fundamental solutions to the
equation $\ddot{\sigma}+\lambda (t)\sigma =0$ and $A$, $B$, $C$ are
constants constrained as $C^{2}=AB-W^{-2}$ with Wronskian $W=u\dot{v}-v\dot{u%
}$. The solutions to the equation with an added dissipative term
proportional to $\dot{\sigma}$ are not known in general. However, the
equation of interest here, (\ref{A1}), which has the linear term removed may
be solved exactly. For this purpose we assume $\varrho (t)$ to be of the form%
\begin{equation}
\varrho (t)=f[q(t)],\qquad \ \ \ \text{with }q(t)=\dint\nolimits^{t}\frac{1}{%
m(s)}ds.  \label{fq}
\end{equation}%
Using this form, equation (\ref{A1}) transforms into%
\begin{equation}
\frac{d^{2}f}{dq^{2}}=\frac{1}{f^{3}},
\end{equation}%
which corresponds to (\ref{EPtime}) with $\lambda (t)=0$. Taking the linear
independent solutions to that equation to be $u(t)=1$ and $v(t)=q$, we obtain%
\begin{equation}
f(q)=\frac{\pm 1}{\sqrt{B}}\sqrt{1+(Bq+C)^{2}}
\end{equation}%
and hence with (\ref{fq}) a solution to (\ref{A1}).

\noindent \textbf{Acknowledgments:} JC and TF are supported by City,
University of London Research Fellowships.

\newif\ifabfull\abfulltrue


\begin{thebibliography}{99}
\bibitem{darboux} G.~Darboux, \newblock On a proposition relative to linear
equations, \newblock physics/9908003, Comptes Rendus Acad. Sci. Paris 
\textbf{94}, 1456--59 (1882).

\bibitem{Witten:1981nf} E.~Witten, \newblock Dynamical breaking of
supersymmetry, \newblock Nucl. Phys. \textbf{B188}, 513 (1981).

\bibitem{Cooper} F.~Cooper, A.~Khare, and U.~Sukhatme, \newblock %
Supersymmetry and quantum mechanics, \newblock Phys. Rept. \textbf{251},
267--385 (1995).

\bibitem{bagrov1} V.~G. Bagrov and B.~F. Samsonov, \newblock Darboux
transformation of the Schr{\"o}dinger equation, \newblock Physics of
Particles and Nuclei \textbf{28}(4), 474 (1997).

\bibitem{MatLongo} S.~Longhi and G.~Della~Valle, \newblock Invisible defects
in complex crystals, \newblock Annals of Physics \textbf{334}, 35--46 (2013).

\bibitem{correa2015p} F.~Correa, V.~Jakubsk{\`y}, and M.~S. Plyushchay, %
\newblock PT-symmetric invisible defects and confluent Darboux-Crum
transformations, \newblock Physical Review A \textbf{92}(2), 023839 (2015).

\bibitem{matveevdarboux} V.~B. Matveev and M.~A. Salle, \newblock Darboux
transformation and solitons, \newblock (Springer, Berlin) (1991).

\bibitem{correahidden} F.~Correa and M.~S. Plyushchay, \newblock Hidden
supersymmetry in quantum bosonic systems, \newblock Annals of Physics 
\textbf{322}(10), 2493--2500 (2007).

\bibitem{CCFsineG} J.~Cen, F.~Correa, and A.~Fring, \newblock Degenerate
multi-solitons in the sine-Gordon equation, \newblock J. Phys. A: Math.
Theor. \textbf{50}, 435201 (2017).

\bibitem{guilarte2017perfectly} J.~M. Guilarte and M.~S. Plyushchay, %
\newblock Perfectly invisible PT-symmetric zero-gap systems, conformal field
theoretical kinks, and exotic nonlinear supersymmetry, \newblock J. of High
Energy Phys. \textbf{2017}(12), 61 (2017).

\bibitem{cen2018asymptotic} J.~Cen and A.~Fring, \newblock Asymptotic and
scattering behaviour for degenerate multi-solitons in the Hirota equation, %
\newblock arXiv preprint arXiv:1804.02013 (2018).

\bibitem{bagrov2} V.~G. Bagrov and B.~F. Samsonov, \newblock Supersymmetry
of a nonstationary Schr{\"o}dinger equation, \newblock Physics Letters A 
\textbf{210}(1-2), 60--64 (1996).

\bibitem{song2003} D.-Y. Song and J.~R. Klauder, \newblock Generalization of
the Darboux transformation and generalized harmonic oscillators, \newblock %
J. of Phys. A: Math. and Gen. \textbf{36}(32), 8673 (2003).

\bibitem{suzko2009darboux} A.~A. Suzko and A.~Schulze-Halberg, \newblock %
Darboux transformations and supersymmetry for the generalized Schr{\"o}%
dinger equations in (1+ 1) dimensions, \newblock J. of Phys. A: Math. and
Theor. \textbf{42}(29), 295203 (2009).

\bibitem{Urubu} F.~G. Scholtz, H.~B. Geyer, and F.~Hahne, \newblock %
Quasi-Hermitian Operators in Quantum Mechanics and the Variational
Principle, \newblock Ann. Phys. \textbf{213}, 74--101 (1992).

\bibitem{Benderrev} C.~M. Bender, \newblock Making sense of non-Hermitian
Hamiltonians, \newblock Rept. Prog. Phys. \textbf{70}, 947--1018 (2007).

\bibitem{Alirev} A.~Mostafazadeh, \newblock Pseudo-Hermitian Representation
of Quantum Mechanics, \newblock Int. J. Geom. Meth. Mod. Phys. \textbf{7},
1191--1306 (2010).

\bibitem{CA} C.~Figueira~de Morisson~Faria and A.~Fring, \newblock Time
evolution of non-Hermitian Hamiltonian systems, \newblock J. Phys. \textbf{%
A39}, 9269--9289 (2006).

\bibitem{time1} A.~Mostafazadeh, \newblock Time-dependent pseudo-Hermitian
Hamiltonians defining a unitary quantum system and uniqueness of the metric
operator, \newblock Physics Letters B \textbf{650}(2), 208--212 (2007).

\bibitem{time6} M.~Znojil, \newblock Time-dependent version of
crypto-Hermitian quantum theory, \newblock Physical Review D \textbf{78}(8),
085003 (2008).

\bibitem{time7} J.~Gong and Q.-H. Wang, \newblock Time-dependent
PT-symmetric quantum mechanics, \newblock J. Phys. A: Math. and Theor. 
\textbf{46}(48), 485302 (2013).

\bibitem{fringmoussa} A.~Fring and M.~H.~Y. Moussa, \newblock Unitary
quantum evolution for time-dependent quasi-Hermitian systems with
nonobservable Hamiltonians, \newblock Physical Review A \textbf{93}(4),
042114 (2016).

\bibitem{AndTom1} A.~Fring and T.~Frith, \newblock Exact analytical
solutions for time-dependent Hermitian Hamiltonian systems from static
unobservable non-Hermitian Hamiltonians, \newblock Phys. Rev. A \textbf{95},
010102(R) (2017).

\bibitem{AndTom2} A.~Fring and T.~Frith, \newblock Metric versus observable
operator representation, higher spin models, \newblock Eur. Phys. J. Plus,
133: 57 (2018).

\bibitem{AndTom3} A.~Fring and T.~Frith, \newblock Mending the broken
PT-regime via an explicit time-dependent Dyson map, \newblock Phys. Lett. A,
2318 (2017).

\bibitem{AndTom4} A.~Fring and T.~Frith, \newblock Solvable two-dimensional
time-dependent non-Hermitian quantum systems with infinite dimensional
Hilbert space in the broken PT-regime, \newblock J. of Phys. A: Math. and
Theor. \textbf{51}(26), 265301 (2018).

\bibitem{mostafazadeh2018energy} A.~Mostafazadeh, \newblock Energy
Observable for a Quantum System with a Dynamical Hilbert Space and a Global
Geometric Extension of Quantum Theory, \newblock arXiv preprint
arXiv:1803.04175 (2018).

\bibitem{AndTom5} A.~Fring and T.~Frith, \newblock Quasi-exactly solvable
quantum systems with explicitly time-dependent Hamiltonians, \newblock %
preprint arXiv:1808.03547, Phys. Lett. A (2018)
https://doi.org/10.1016/j.physleta.2018.10.043.

\bibitem{crum} M.~M. Crum, \newblock Associated Sturm-Liouville systems, %
\newblock The Quarterly Journal of Mathematics \textbf{6}(1), 121--127
(1955).

\bibitem{gelfand2005} I.~Gelfand, S.~Gelfand, V.~Retakh, and R.~L. Wilson, %
\newblock Quasideterminants, \newblock Adv. Math \textbf{193}(1), 56--141
(2005).

\bibitem{CorreaFring} F.~Correa and A.~Fring, \newblock Regularized
degenerate multi-solitons, \newblock Journal of High Energy Physics \textbf{%
2016}(9), 8 (2016).

\bibitem{CenFringHir} J.~Cen, F.~Correa, and A.~Fring, \newblock Integrable
nonlocal Hirota equations, \newblock arXiv:1710.11560 (2017).

\bibitem{maamache2017pseudo} M.~Maamache, O.~K. Djeghiour, N.~Mana, and
W.~Koussa, \newblock Pseudo-invariants theory and real phases for systems
with non-Hermitian time-dependent Hamiltonians, \newblock The European
Physical Journal Plus \textbf{132}(9), 383 (2017).

\bibitem{Lewis69} H.~Lewis and W.~Riesenfeld, \newblock An Exact quantum
theory of the time dependent harmonic oscillator and of a charged particle
time dependent electromagnetic field, \newblock J. Math. Phys. \textbf{10},
1458--1473 (1969).

\bibitem{GV1} W.~Gordon, \newblock Der Comptoneffekt nach der Schr{\"o}%
dinger Theorie, \newblock Zeit. f{\"u}r Physik \textbf{40}, 117--133 (1926).

\bibitem{GV2} D.~M. Volkov, \newblock On a class of solutions of the Dirac
equation, \newblock Zeit. f{\"u}r Physik \textbf{94}, 250 (1935).

\bibitem{Rev1} C.~Figueira~de Morisson~Faria, A.~Fring, and R.~Schrader, %
\newblock Analytical treatment of stabilization, \newblock Laser Physics 
\textbf{9}, 379--387 (1999).

\bibitem{plasmonic} W.~Liu, D.~N. Neshev, I.~V. Shadrivov, A.~E.
Miroshnichenko, and Y.~S. Kivshar, \newblock Plasmonic Airy beam
manipulation in linear optical potentials, \newblock Optics Letters \textbf{%
36}(7), 1164--1166 (2011).

\bibitem{miller1977symmetry} W.~Miller~Jr, \newblock Symmetry and separation
of variables, \newblock (1977).

\bibitem{berry1979} M.~V. Berry and N.~L. Balazs, \newblock Nonspreading
wave packets, \newblock Am. J. of Phys. \textbf{47}(3), 264--267 (1979).

\bibitem{gori1999general} F.~Gori, M.~Santarsiero, R.~Borghi, and
G.~Guattari, \newblock The general wavefunction for a particle under uniform
force, \newblock Euro. J. of Phys. \textbf{20}(6), 477 (1999).

\bibitem{siviloglou2007o} G.~A. Siviloglou, J.~Broky, A.~Dogariu, and D.~N.
Christodoulides, \newblock Observation of accelerating Airy beams, \newblock %
Phys. Rev. Lett. \textbf{99}(21), 213901 (2007).

\bibitem{siviloglou2007a} G.~A. Siviloglou and D.~N. Christodoulides, %
\newblock Accelerating finite energy Airy beams, \newblock Optics Lett. 
\textbf{32}(8), 979--981 (2007).

\bibitem{baumgartl2008} J.~Baumgartl, M.~Mazilu, and K.~Dholakia, \newblock %
Optically mediated particle clearing using Airy wavepackets, \newblock %
Nature photonics \textbf{2}(11), 675 (2008).

\bibitem{vettenburg2014light} T.~Vettenburg, H.~I.~C. Dalgarno, J.~Nylk,
C.~Coll-Llad{\'o}, D.~E.~K. Ferrier, T.~{\v{C}}i{\v{z}}m{\'a}r, F.~J.
Gunn-Moore, and K.~Dholakia, \newblock Light-sheet microscopy using an Airy
beam, \newblock Nature methods \textbf{11}(5), 541 (2014).

\bibitem{patsyk2018} A.~Patsyk, M.~A. Bandres, R.~Bekenstein, and M.~Segev, %
\newblock Observation of Accelerating Wave Packets in Curved Space, %
\newblock Phys. Rev. X \textbf{8}(1), 011001 (2018).

\bibitem{Swanson} M.~S. Swanson, \newblock Transition elements for a
non-Hermitian quadratic Hamiltonian, \newblock J. Math. Phys. \textbf{45},
585--601 (2004).

\bibitem{masst} M.~Davidson, \newblock Variable mass theories in
relativistic quantum mechanics as an explanation for anomalous low energy
nuclear phenomena, \newblock in \emph{J. of Phys.: Conference Series},
volume 615, page 012016, IOP Publishing, 2015.

\bibitem{pedrosa1997exact} I.~A. Pedrosa, \newblock Exact wave functions of
a harmonic oscillator with time-dependent mass and frequency, \newblock %
Phys. Rev. A \textbf{55}(4), 3219 (1997).

\bibitem{Ermakov} V.~Ermakov, \newblock Transformation of differential
equations,, \newblock Univ. Izv. Kiev. \textbf{20}, 1--19 (1880).

\bibitem{Pinney} E.~Pinney, \newblock The nonlinear differential equation $%
y^{\prime \prime 3}=0$, \newblock Proc. Amer. Math. Soc. \textbf{1}, 681(1)
(1950).
\end{thebibliography}
\end{document}